\documentclass[11pt,a4paper]{article}
\usepackage{amsmath}
\usepackage{amsthm}
\usepackage{amssymb}
\usepackage{natbib}
\usepackage{siunitx}
\usepackage{amsfonts}
\usepackage[margin=3cm]{geometry}
\usepackage{parskip}
\usepackage[T1]{fontenc}
\usepackage{subcaption}
\usepackage{floatrow}
\usepackage{setspace}
\usepackage{hyperref}
\usepackage{graphicx}
\usepackage{titlesec}
\usepackage[svgnames]{xcolor}
\DeclareMathOperator*{\argmax}{arg\,max}

\counterwithin{figure}{section}
\counterwithin{table}{section}
\hypersetup{colorlinks=true,linkcolor=Crimson,citecolor=Crimson,urlcolor=Crimson}
\newcommand{\mR}[0]{\mathbb{R}}
\newcommand{\mE}[0]{\mathbb{E}}
\newcommand{\QuoteExplorationOptimized}{\emph{Soft} actor-critic algorithms optimise exploration by imposing a constraint on the Kullback-Leibler divergence\footnote{Kullback-Leibler divergence measures non-commutatively how different two probability distributions are from each other, and is defined for two continuous densities $f(\cdot), g(\cdot)$ over support $\Omega$ as $\int_\Omega f(x) \log [f(x) / g(x)] dx$.} between the policy and a uniform distribution, that is, they learn the policy that maximises long-run profit while keeping enough randomness to continue learning \citep{neu_unified_2017}.
Exploration is thus optimised instead of uniform.
The algorithm can learn low variance near high-reward, low-loss states and higher variance in states where the estimated reaction function is less certain, as measured by higher errors in the estimation of the action-value function.}

\newcommand{\QuoteAlgorithmAxes}{The algorithm I employ, soft actor-critic \citep{haarnoja_soft_2018}, differs from tabular Q-learning in two main regards, which constitute two of the main degrees of freedom\footnote{Soft actor-critic and Q-learning are both off-policy algorithms, in the sense that they both aim to optimise the average reward of the \emph{optimal} policy and compare observed rewards to the potential ``optimal stream''.
This is in contrast to on-policy algorithms, which optimise the average reward of the \emph{current} policy.
On-policy algorithms tend to be faster in terms of wall-clock time per iteration and simpler to implement, but fare worse when data collection is expensive, like in algorithmic pricing, because they require significantly more data to learn a good policy.}
in algorithm design in reinforcement learning research (see chapters 9 to 12 in \citealt{sutton_reinforcement_2018}).
First, it is a \emph{function approximation} approach, that is, it maintains parametric estimates of both the action-value function $q(s, a; \boldsymbol{w})$, or critic, and the optimal policy $\sigma(a|s; \boldsymbol{\theta})$, or actor, where $\boldsymbol{\theta}$ and $\boldsymbol{w}$ are parameter vectors.
Second, it is a \emph{policy-gradient} algorithm, that is, it learns its policy $\sigma(\cdot|\cdot)$ as a continuous, parametric function of the learned action-value function $q(\cdot)$.
Policy gradient algorithms are the current state of the art in reinforcement learning research, and have been shown to be very effective in complex environments, from robotics to autonomous driving \citep{kiran_deep_2021}.
The choice of a function approximation, policy gradient approach affects (i) flexibility over the choice of action space (ii) the formulation of the optimisation problem to be solved, and (iii) the direction of exploration.
In the rest of this section, I will explain these consequences of algorithmic design.
}

\newcommand{\QuotePhaseDiagramMethod}{I thus generate the following phase diagrams by taking a ``snapshot'', or ``checkpoint'', of the policy network at the end of the simulation, and plotting the resulting price trajectories by iteratively applying the policy.
Phase diagrams generated this way only depend on the period $t$ at which the checkpoint is saved and on the initial state $p_1, p_2$.
Indeed, the status of the replay buffer, the critic network, or the policy variance at $t$ have no bearing on the mean policy at $t$: it is only a function of the actor network parameters at $t$.}

\newcommand{\QuotePolicyResponse}{The primary pattern visible is the global stability of streamlines towards fixed points.
Deviations from the fixed point are met with immediate corrective adjustments that drive the system back towards the equilibrium.
These adjustments do not uniformly flow toward the fixed point, but first curve perpendicularly to the direction of a deviation from the fixed point, indicating a responsive, ``tit-for-tat'' style strategy: each agent sets low prices when the adversary is posting low prices and high prices when the adversary is setting high prices.
The high fixed point in this representative run is situated significantly higher than the competitive floor, showing a learned supracompetitive pricing rule rather than an unconditional high-price rule.}

\newcommand{\QuoteSequentialExtension}{A natural extension would be to move from simultaneous-move pricing to sequential pricing environments as in \citet{klein_autonomous_2021}, to study how the presence of a price leader affects the emergence and stability of algorithmic collusion.
Preliminary results available upon request suggest that in this case, play with mutual best responses is more prevalent, profit gains are higher, and the timescale of collusion is similar.}

\newcommand{\QuoteRtwoAbstract}{Artificial intelligence algorithms are increasingly used by firms to set prices.
Previous research shows that they can learn to collude, but how quickly and reliably they can do so has remained an open question.
I study average-reward soft actor-critic algorithms that set continuous prices in the workhorse environment of this literature: a repeated Bertrand duopoly with symmetric firms, deterministic logit demand, and constant marginal costs, where agents observe only market prices and their own profits.
The algorithms can converge to a collusive outcome within 50{,}000 periods, and in 36\% of simulations learned punishments make one-period deviations unprofitable, so that high prices are mutual best responses.
Successful simulations show cooperative behaviour supported by reward-punishment schemes that discourage deviations.
When high prices are not supported by punishments, profitable deviations are typically on the order of hundredths of a percent.}

\newcommand{\QuoteRtwoIntroContribution}{I analyse a class of algorithms more advanced than Q-learning and demonstrate that, in a subset of runs in the same stylised environment, these algorithms can autonomously learn collusive pricing with mutual best responses much faster than \citet{calvano_artificial_2020}, reducing the horizon from about one million periods to about 50{,}000.}

\newcommand{\QuoteRtwoAverageRewardIntro}{The average-reward formulation is important in pricing applications because it avoids choosing an exogenous discount factor for short pricing periods; I discuss this point in detail in the methodology section and Appendix A.}

\newcommand{\QuoteRtwoPracticalLimitation}{Deep reinforcement learning algorithms are notoriously unstable, with around two-thirds of trials routinely failing to learn any sensible control rule in continuous control tasks \citep{henderson_deep_2018}.
On the other hand, in 36\% of tested agent-level deviations in the baseline duopoly simulations, two soft actor-critic algorithms competing à la Bertrand maintain a supra-competitive fixed point with mutual best responses in around 50{,}000 periods under logit, non-stochastic demand and constant marginal costs, no capacity constraints, no entry or exit, and no informational asymmetry.
If a period is one hour, this would be about six years.
The model preserves the stylised environment of previous proof-of-concept work, and the result concerns whether more modern learning algorithms reduce the simulated convergence time within that environment.}

\newcommand{\QuoteRtwoQMatrix}{For instance, in the baseline analysis of \citet{calvano_artificial_2020}, the Q-matrix comprises over 3,000 cells\footnote{There are 225 states, given by 15 actions per agent, and thus $225 \times 15 = 3{,}375$ state-action pairs.}, so visiting each Q-table cell 20 times on average, under uniform permanent exploration, would require $3{,}375 \times 20 = 67{,}500$ total periods.}

\newcommand{\QuoteRtwoNoGuarantee}{On the one hand, in this multi-agent pricing environment there is no general guarantee that any reinforcement learning agent, tabular or function-approximation-based, converges to a Nash equilibrium or even to a stable useful policy.}

\newcommand{\QuoteRtwoExploration}{Finally, soft actor-critic represents the strategy profile by a parameterised policy that has to be differentiable, or at least subdifferentiable, with respect to the actor-network parameters, because the policy is updated by stochastic gradient descent.
This differentiable policy is stochastic, mapping states to a price distribution rather than directly to a single price, so exploration is part of the strategy profile being optimised rather than a separate ``exploration-exploitation'' switch.}

\newcommand{\QuoteRtwoFailureLearning}{High prices may be the result of simply failing to learn the optimal strategy.
For example, parameter updates may assign high mean prices to a wide region of the state space because high prices are profitable on the realised path, without learning a conditional response that punishes deviations.}

\newcommand{\QuoteRtwoOvertaking}{Any finite transient has zero weight in the average-overtaking limit because it is divided by $T$ as $T \to \infty$.}

\newcommand{\QuoteRtwoConvergenceCriterion}{\citet{calvano_artificial_2020} define convergence by checking whether, for every firm and every state on the finite price grid, the learned optimal action remains unchanged for 100{,}000 consecutive periods.
In the present continuous stochastic-policy setting, one could analogously monitor changes in the mean policy evaluated on a grid, or more coarsely monitor first differences or standard deviations of played prices.
However, algorithms may still be optimising their strategies while prices are stable, so evaluating performance as soon as prices stabilise would not be representative of the actual behaviour of the algorithms.
In general, it would be up to the firm to decide when the algorithm has achieved satisfactory performance and can be deployed without exploration.
Therefore, I let simulations run for a fixed amount of time steps, and study the behaviour of the time series of prices played and their limit strategies, that is, the \emph{mean} strategies they play in the last 5{,}000 periods of the simulation.
}

\newcommand{\QuoteRtwoDeterministicEvaluation}{During training, each soft actor-critic agent plays mixed strategies, sampling prices from a state-dependent Gaussian policy.
For evaluation, I make them play pure strategies instead.
I freeze the actor-network weights and use the policy mean, equivalently setting the policy variance to zero for the evaluated path; Appendix B gives the details.
In this deterministic evaluation of end-of-simulation policies, no sessions converged to price cycles.}

\newcommand{\QuoteRtwoSettlingWindow}{The 50-period window is a settling window: it lets deterministic dynamics reach their local fixed point from the last realised training state and is longer than the return paths shown below.}

\newcommand{\QuoteRtwoHistogram}{The bars in the figure are histogram bins; the central bar represents small gains close to zero.}

\newcommand{\QuoteRtwoGamma}{To enable comparisons with the Q-learning of \citet{calvano_artificial_2020}, I use an evaluation discount factor $\delta = 0.95$ in evaluating discounted profit.}

\newcommand{\QuoteRtwoFixedPoints}{One fixed point is at the Bertrand-Nash benchmark.
The other is above the cooperative-price benchmark because of simulation noise.}

\newcommand{\QuoteRtwoThreeAgents}{For comparability with the two-agent result above, I focus on the final 5{,}000 observations of each run rather than average profits that include the initial exploration and low-price learning phase.
Mean profit gain in this final window is about 30\% with two agents and about 25\% with three agents, as shown by Figure 3.7, but sessions with much higher profit gains remain.}

\newcommand{\QuoteRtwoDiscussionCaveat}{The evidence should nevertheless be interpreted conservatively.
The baseline environment keeps the restrictive assumptions of the proof-of-concept literature: symmetric firms, deterministic logit demand, constant marginal costs, no capacity constraints, no entry or exit, and no informational asymmetry.
I show that an important bottleneck in the earlier algorithmic collusion literature, that is, the speed of learning, can be loosened without injecting explicit economic structure that would be detrimental to the definition of ``autonomous'' collusion.}

\newcommand{\QuoteRtwoMultiPeriod}{The reported diagnostics evaluate a one-period static-best-response deviation followed by a return to the learned policy.
As a robustness check, I verified that, at the end of training, 83\% of deviations that are unprofitable in one period remain unprofitable when repeated for five periods, and that the profitable exceptions gain amounts on the order of hundredths of a percent\footnote{The five-period deviations repeat the static-best-response price computed at the first deviation period. Detailed results of this check are available upon request.}.
I also checked that the static best response yields, on average, weakly higher discounted profit than any other tested deviation\footnote{Other deviations have been evaluated by applying the same procedure to a discrete, equally-spaced grid of prices to deviate to. Detailed results of this check are available upon request.}, that is, if deviating to the static best response to the last played price is not profitable, deviating to any other tested price is, on average, not profitable either.}

\begin{document}
\title{Convergence to collusion in algorithmic pricing\thanks{I sincerely wish to thank Emilio Calvano, Giacomo Calzolari, Vincenzo Denicolò, Massimiliano Furlan, Santiago Ortiz, Xavier Lambin, Sergio Pastorello, Violetta van Veen, two anonymous referees, and Editor Özlem Bedre Defolie for precious comments and suggestions concerning all drafts of this article.
The author acknowledges financial support from EUR CHESS (ANR 17 EURE 0010).
Any remaining errors are the author's sole responsibility.
Replication code, based on a framework by \citet{hettich_algorithmic_2021}, is available under the GNU Affero General Public License v3.0 at \url{https://github.com/kmfrick/price_simulator}.
This paper supersedes previous drafts titled ``Algorithmic pricing using policy gradient reinforcement learning'' and ``Convergence rates and collusive outcomes of pricing algorithms''.
The author declares no conflicts of interest.
}}
\author{Kevin Michael Frick\thanks{Toulouse School of Economics, University of Toulouse Capitole, Toulouse, France, \href{mailto:kevin-michael.frick@tse-fr.eu}{\texttt{kevin-michael.frick@tse-fr.eu}}}}
\date{\today}
\maketitle
\begin{abstract}
\QuoteRtwoAbstract
\end{abstract}

\clearpage

\section{Introduction}

In recent years, legislative attention highlights a serious threat to the antitrust policy framework: the potential for artificial intelligence to undermine market competition \citep{oecd_algorithms_2017,calvano_protecting_2020,uscongress_preventing_2024}.
A prominent body of literature in economics and computer science has thus emerged to investigate whether pricing algorithms can indeed autonomously learn collusive behaviour.

The possibility of algorithmic collusion was demonstrated by \citet{calvano_artificial_2020} and \citet{klein_autonomous_2021} using pricing algorithms based on Q-learning.
Their findings have been replicated and expanded \citep{asker_artificial_2022,banchio_adaptive_2022,johnson_platform_2023,calvano_algorithmic_2023,abada_collusion_2024}.
Moreover, empirical evidence of algorithmic collusion has emerged.
An important contribution by \citet{assad_algorithmic_2023} analysed the impact of widespread adoption of algorithmic pricing in the German retail gasoline market.
The study found that nearby competitors' adoption of pricing algorithms increased their mean margins by almost a third compared to pre-adoption levels, whereas sole adoption did not lead to significant margin changes.

The empirical analysis in \citet{assad_algorithmic_2023} aligns with \citet{calvano_artificial_2020} by demonstrating a gradual increase in margins over time, suggesting that algorithms have a learning curve. However, the time scales differ noticeably. \citet{assad_algorithmic_2023} revealed a collusion timescale of a couple of years, consistent with the findings of \citet{byrne_learning_2019} in the case of human collusion.
In contrast, \citet{calvano_artificial_2020} and subsequent experimental literature suggest that convergence to collusive outcomes takes a large number of periods, in the order of hundreds of thousands.
The actual length of a period may vary across industries, but it must be long enough for new prices' demand effects to materialise and observable profits to emerge.
In \citet{assad_algorithmic_2023}, prices change on average every hour, making the one million time steps required for convergence in \citet{calvano_artificial_2020} equivalent to dozens of years of real-life time.
Such discrepancies may challenge the applicability of experimental and theoretical results and the solidity of the empirical evidence \citep{schwalbe_algorithms_2018,den_artificial_2022}.

Acknowledging the problem of the slow learning time scale, \citet{calvano_artificial_2020} attribute it to the nature of Q-learning algorithms, which learn slowly by design.
They speculate that more sophisticated algorithms might learn to collude more quickly.
The primary contribution of this article is to verify this conjecture.
\QuoteRtwoIntroContribution

The slow learning rate of Q-learning algorithms is due to their estimation of a Q-matrix, containing the value of each possible action (typically, a firm's price) in each possible state of the dynamical system (typically, a vector of prices in past periods).
This estimation starts from arbitrary initial values and updates the matrix based on observations of the firm's profits for each possible action-state combination. This approach has several implications.
First, the action space, which would normally be continuous, must be discretised.
Second, learning requires visiting all cells of the matrix, necessitating random exploration instead of always choosing the action deemed optimal based on acquired knowledge.
Third, the updating must be gradual to prevent rapid forgetting of past learning, requiring each cell to be visited multiple times.
\QuoteRtwoQMatrix
However, collusion cannot be learned during exploration mode because the best response to uniform random pricing by a rival cannot be a collusive strategy, as there is no scope for punishment for deviation from collusion.
Consequently, the algorithms must be in exploitation mode frequently enough, leading to the need for approximately one million periods for convergence.

To overcome this issue, this article employs average-reward soft actor-critic algorithms \citep{haarnoja_soft_2018,adamczyk_average_2025}, the current state of the art in the reinforcement learning literature.
\QuoteRtwoAverageRewardIntro

An \emph{actor-critic} algorithm learns to estimate a continuous action-value function rather than a discrete matrix, and at the same time learns to estimate a continuous strategy profile that takes as arguments the estimated action values.
These algorithms exploit the topological structure of the action and strategy spaces, which Q-learning disregards, such that after observing the profit for a particular combination of current and past prices, the algorithms update the value not only for that specific point but also for nearby prices.
Another innovation concerns the direction of exploration.
\QuoteExplorationOptimized

I show that the combined effect of these natural and sensible changes to the learning strategy of Q-learning algorithms is two-fold.
\QuoteRtwoNoGuarantee
\QuoteRtwoPracticalLimitation

The rest of this article is organised as follows.
Section \ref{chap:method} describes the model I employ and the economic environment in which I study its behaviour.
Section \ref{chap:results} details how agents behave, how their behaviour changes during learning, and the strategies they employ to maintain cooperative behaviour.
Section \ref{chap:discussion} places the results in the broader context of the critical literature on reinforcement learning in pricing games.
Section \ref{chap:conclusions} concludes.

\section{Methodology}
\label{chap:method}

In this section I specify a policy gradient pricing algorithm, drawing from state-of-the-art results in the reinforcement learning literature, and the economic model to study its behaviour.

\subsection{Reinforcement learning}

Reinforcement learning algorithms are concerned with learning a policy, or strategy profile, that maximises total reward in a Markov decision process defined (\citealt{van_hasselt_reinforcement_2012}, sec. 1.1) by a set of states $\mathcal{S}$, a set of actions $\mathcal{A}$, a reward function\footnote{In the reinforcement learning literature, dependence of the reward function on future states allows the reward process to be stochastic and codetermined with the state transition process, e.g. \citet{van_hasselt_reinforcement_2012} defines the reward function as potentially depending on $S_{t+1}$.
In the specific context of this article, competitors' prices in period $t$ will enter $S_{t+1}$ but also determine demand and thus rewards in period $t$.
To ease notation, I let this relation with other players' actions be absorbed by $\pi$ and drop dependence on $S_{t+1}$ from the notation.} $\pi: \mathcal{S} \times \mathcal{A} \rightarrow \mR$ and a state transition probability density function $T(s_{t+1}|a_t, s_t, s_{t - 1}, ..., s_0) \equiv \Pr(S_{t+1} = s |A_t, S_t, S_{t-1}, ..., S_0) $ that satisfies the Markov assumption: $T(s_{t+1} |a_t, s_t, s_{t - 1}, ..., s_0) = T(s_{t + 1} | a_t, s_t)$.
A strategy profile $\sigma: \mathcal{S} \times \mathcal{A} \rightarrow \mR$ is a probability density function, conditional on the current state $S_t = s \in \mathcal{S}$, whose support is the action space $\mathcal{A}$.

The objective of these algorithms is to learn the optimal strategy profile $\sigma^*(a|s)$, that is the one that maximises the expected present value of the reward process:
\begin{equation}
  \mE_\sigma \left[\sum_{t=0}^\infty \delta^t \pi(S_t, A_t)\right]
\end{equation}
where $\delta$ is the discount factor and the $\mE_\sigma[\cdot]$ operator denotes an expectation with respect to the state transition function of the Markov decision process and conditional on the agent following strategy profile $\sigma(a|s)$.

The optimal strategy profile is the one that maximises the value and action-value function for all possible states.
Most reinforcement learning algorithms approach this optimisation problem by estimating the action-value function $q_\sigma: \mathcal{S} \times \mathcal{A} \rightarrow \mR$, a precursor of Bellman's value function $v_\sigma: \mathcal{S} \rightarrow \mR$.
The action-value function gives the expected total discounted reward that the agent receives starting in state $s$, taking action $a$, and following strategy profile $\sigma$ from that point on:
\begin{equation}
  q_\sigma (s, a) \equiv \mE_\sigma\left[\pi(S_t, A_t) + \delta v_\sigma(S_{t+1}) \middle| S_t = s, A_t = a\right] \quad \forall \; (s, a) \in \mathcal{S} \times \mathcal{A}.
  \label{eq:qdisc}
\end{equation}
Bellman's value function is given by $v_\sigma(s) = \mE_\sigma[q_\sigma(S_t, A_t) | S_t = s]$.
If the agent knew $q_\sigma(s,a)$ for any given strategy, the problem would be solved by following the strategy that maximises action-values for any state and action and always selects the action with the highest action-value in any state, i.e. $q^*(s, a)$ solves the Bellman equation
\begin{equation}
  q^*(s, a) = \max\limits_{\sigma} \mE_\sigma[\pi(S_t, A_t) + \delta \max_{a' \in \mathcal{A}} q_\sigma(S_{t+1}, a') | S_t = s, A_t = a]
  \label{eq:bellmanq}
\end{equation}
and $\sigma^*(a|s) = \mathbf{1}\left[a = \argmax\limits_{\hat{a} \in \mathcal{A}} q^*(s, \hat{a})\right]$ is a degenerate probability function that assigns probability one to the optimal action and zero to all other actions.

When the action-value function is not known, it can be estimated.
The Q-learning algorithm, currently the standard in the algorithmic pricing literature \citep{calvano_artificial_2020,klein_autonomous_2021,asker_artificial_2022,johnson_platform_2023}, maintains a matrix representation $\boldsymbol{Q} \in \mR^{|\mathcal{S}| \times |\mathcal{A}|}$ of the action-value function, known as the Q-matrix, and updates one cell at every iteration.
This approach quickly becomes unfeasible as state and action spaces grow and is inapplicable to continuous state and/or action spaces.
Most importantly, updating only one state-action combination at a time\footnote{As $q_\sigma(s, a)$ is an ordinary function, it can be estimated using function approximation approaches (\citealt{sutton_reinforcement_2018}, chap. 9) such as tile coding or an artificial neural network.
To the best of my knowledge, the only demonstration of a possibility of improving convergence times using function approximation to estimate the action-value function is due to \citet{hettich_algorithmic_2021}.
He shows that it is possible to observe faster convergence in the economic environment in \citet{calvano_artificial_2020} by employing deep Q-learning, an evolution of Q-learning \citep{mnih_human-level_2015} that uses a neural network to estimate $q(s, a)$.
That said, the improvement provided by such methods is still not sufficient to justify the emergence of collusion on a commercially plausible timescale, and they still share the limitations of Q-learning such as a discrete action set.} means that it needs a large amount of experience to converge to the optimal strategy profile.
\citet{asker_artificial_2022} term algorithms such as Q-learning ``asynchronous'', in contrast with ``synchronous'' learning which uses information from economic theory to construct counterfactual profits and update multiple points of the value function estimation at a time.
In this sense, actor-critic algorithms are closer to ``synchronous'' learning, as changing even one parameter of the model amounts to changing the action-value function estimation for all state-action pairs, and as opposed to Q-learning, they incorporate knowledge from batches of observations rather than just the latest one (see Appendix \ref{appendix:specs} for details).
However, updates are not driven by injected knowledge about economic theory -- learning entirely relies on observed experience.

The issue of the speed of convergence in reinforcement learning is a long-standing one in the computer science literature \citep{cesa_prediction_2006,yu_towards_2018,foster_statistical_2021,liu_sample_2022}, where convergence speed is usually termed \emph{sample efficiency} or sample complexity, by analogy with computational complexity.
\citet{yu_towards_2018} argues that the exploration strategy, the optimisation process and modelling choices on state and action spaces are the main determinants of sample efficiency.
Informed by this literature, in the following section I discuss how modern reinforcement learning algorithms address these aspects and their relevance for the algorithmic pricing problem.

\subsection{Specification of the algorithm}

\QuoteAlgorithmAxes

Soft actor-critic uses artificial neural networks as function approximators.
The actor network is trained to choose the action that maximises the action-value function as computed by the critic network (\citealt{haarnoja_soft_2018}, eq. 1).
The critic network is trained to minimise the difference between the action-value function and ``observed action values'' that are computed on the basis of the observed profit stream (\citealt{haarnoja_soft_2018}, eq. 5).
I report technical details on the algorithm in Appendix \ref{appendix:specs}.

Policy gradient algorithms can operate on a continuous price space (\citealt{van_hasselt_reinforcement_2012}, sec. 3.2.1).
By doing so, the optimisation process can exploit the topological structure of the action and state spaces.
Q-learning represents actions as belonging to an unordered set (\citealt{sutton_reinforcement_2018}, eq. 3.2) so it has no notion of which action represents a higher or a lower price, and the ordering of actions is merely for human convenience and interpretation.
Conversely, estimating the strategy profile function by training a neural network with gradient descent allows for a more well-defined representation of the algorithm's task as a constrained optimisation problem solved using a stochastic gradient descent routine that maximises a well-specified objective function and is aware of the ordering of actions and states (\citealt{sutton_reinforcement_2018}, chap. 13).
The resulting strategy profile is a function of past prices whose support is the exogenously specified space of possible prices.
In addition, working on a continuous price space avoids the trade-off faced by Q-learning between minimising discretisation error and keeping the size of the Q-matrix manageable in memory.

In addition to being restricted to operating on a discrete state and action space, Q-learning faces an issue related to the scale of the discount factor.
If we argue that one time step represents a few hours of real time, we get a \emph{yearly} discount factor that is implausibly close to zero\footnote{For example, if we take a discount factor $\delta = 0.99$ and say that half an hour passes for every time step, we get a yearly discount factor in the order of $10^{-77}$. With two hours per time step, the yearly discount factor is in the order of $10^{-20}$.}.
However, we cannot set the discount factor at each time step too close to 1, as numerical issues would arise with very large Q-matrix entries, a well-known phenomenon in the computer science literature.
When using function approximation instead, maximising discounted profit is not a well-defined optimisation problem \citep{naik_discounted_2019}, and reinforcement learning algorithms should aim to maximise average profit $\bar{\pi}(\sigma)$, a different metric that does not include a discount factor and whose maximisation is a well-defined problem under function approximation
\footnote{While \citet{possnig_learning_2023,possnig_reinforcement_2023} shows that certain actor-critic algorithms converge to collusion, his results do not apply to the methods used in this article.
His reliance on maximizing discounted profit makes the problem ill-posed in continuous state spaces \citep{singh_learning_1994,naik_discounted_2019}.
In addition, his proofs do not extend to the overparameterised neural networks and stochastic strategy profiles employed here.}.

$\bar{\pi}(\sigma)$ represents the average profit by state weighted by how much time the agent spends in that state, i.e.
\begin{align}
  \bar{\pi}(\sigma) &\equiv \lim\limits_{t \rightarrow \infty} \mE\left[\pi(S_t, A_t) \middle| (A_0, ..., A_t \sim \sigma(a|s;\boldsymbol{\theta})), S_0\right]
   \\ &=\int_\mathcal{S} \mu_\sigma(s) \int_\mathcal{A}\sigma(a|s;\boldsymbol{\theta}) \int_\mathcal{S}\pi(s, a, s') T(s' | s, a) ds' \, da \, ds
  \label{eq:avgrew}
\end{align}
where state visitation is represented by a strategy-dependent probability density function $\mu_\sigma(s)$.

Under the average-reward maximisation problem, value and action-value functions are redefined in their differential formulation:
\begin{equation}
v_\sigma(s) \equiv \mE_\sigma \left[\pi(S_t, A_t) - \bar{\pi}(\sigma) + v_\sigma(S_{t+1})\middle| S_t = s \right]
  \label{eq:vdiff}
\end{equation}\begin{equation}
q_\sigma(s, a) \equiv \mE_\sigma \left[ \pi(S_t, A_t) - \bar{\pi}(\sigma) + v_\sigma(S_{t+1})\middle| S_t = s, A_t = a \right].
\end{equation}
See Appendix \ref{appendix:avgrew} for details on this formulation and on the convergence state visitation measure $\mu_\sigma(s)$.

Finding the optimal representable strategy profile in the average-reward formulation, also known as the bias-optimal strategy profile, is equivalent to finding the optimal strategy profile according to the more familiar average-overtaking criterion \citep{rubinstein_equilibrium_1979,mahadevan_optimality_1996}.
Under this criterion, the optimal strategy profile $\sigma^*$ is such that steady-state performance eventually beats or matches any other representable strategy profile.
Formally, the value functions will be such that
\begin{equation}
\lim_{T\to \infty} \frac{1}{T} \sum_{t=0}^T [v_{\sigma^*}(S_t) - v_\sigma(S_t)] \geq 0 \; \forall \; \sigma: \mathcal{S} \rightarrow \mR.
\end{equation}
\QuoteRtwoOvertaking

\QuoteRtwoExploration
Reinforcement learning algorithms have to maintain exploration of the action space during the learning process, especially during the initial phases, to avoid being stuck with a sub-optimal strategy profile.
Q-learning can maintain exploration by following the learned strategy profile with some probability $1 - \varepsilon$ and choose a random action instead with probability $\varepsilon$.
This is an exogenously specified learning strategy that is constant across actions and states\footnote{It is possible to make the exploration strategy depend on the current Q-matrix, for example by employing Boltzmann exploration \citep{borgers_learning_1997}, and it is useful in some cases to do so, for example when we require convergence in games with mixed equilibria \citep{kianercy_dynamics_2012}.
However, such a strategy does not necessarily improve the rate of convergence and leads to failure in optimisation unless the learning rate is appropriately adjusted \citep{cesa-bianchi_boltzmann_2017}.}, and is only varied by decaying the value of $\varepsilon$ over time.

Instead, to maintain exploration, soft actor-critic imposes a constraint on the Kullback-Leibler divergence between the learned strategy profile and a uniform distribution.
This constraint is computed as an expectation over all possible values of the conditioning state, so it is possible for the algorithm to learn to explore less in states where there is a clearly superior plan of action (e.g. while actively colluding) and more in states that have been visited less and where the strategy profile may not be optimal yet (e.g. while learning to punish optimally and return to collusion).
Soft actor-critic introduces the exploration constraint by applying the method of Lagrange multipliers to the action-value function, incorporating the constraint as a regularisation term in the reward function to be maximised \citep{dudik_maximum_2006}.
The optimisation routine then tunes which states have to have a lower or higher intensity of exploration.

Formally, the constraint is $h(\sigma(a|s; \boldsymbol{\theta})) \geq \bar{h}$, where $h(\cdot)$ is defined as the Kullback–Leibler divergence with a uniform distribution, minus a support-dependent, constant shift:
\begin{equation}
h(\sigma(a|s; \boldsymbol{\theta})) = - \int_\mathcal{S} \mu_\sigma(s) \int_\mathcal{A} \sigma(a|s;\boldsymbol{\theta}) \log \sigma(a|s;\boldsymbol{\theta}) da \, ds = - \mE_\sigma [ \log \sigma(a|s;\boldsymbol{\theta})].
\label{eq:diffent}
\end{equation}
The constraint enters the action-value function as
\begin{equation}
  q_{\sigma,soft}(s, a) = \mE_\sigma \left[\pi(S_t, A_t) - \bar{\pi}(\sigma) - \alpha \log \sigma(A_t|S_t) + v_{\sigma, soft}(S_{t+1}) \middle| S_t = s, A_t = a\right]
\end{equation}
where $v_{\sigma, soft}(s) = \mE_\sigma[q_{\sigma, soft}(S_t,A_t) | S_t = s]$.

Therefore, the actor's optimisation problem is
\begin{equation}
  \max\limits_{\boldsymbol{\theta}} \mE_{\sigma(\boldsymbol{\theta})} \left[q(s, a; \boldsymbol{w}) - \alpha \log \sigma(a|s; \boldsymbol{\theta})\right]
  \label{eq:sacpiloss}
\end{equation}
and the critic's is
\begin{equation}
  \min\limits_{\boldsymbol{w}} \mE_{\sigma(\boldsymbol{\theta})} \left[q(s, a; \boldsymbol{w}) - \left(\underbrace{\pi(s, a, s') - \bar{\pi}(\sigma(\boldsymbol{\theta})) - \alpha \log \sigma(a|s; \boldsymbol{\theta}) + v_{\sigma(\boldsymbol{\theta}), soft}(s'; \boldsymbol{w})}_{q_{\sigma(\boldsymbol{\theta}),soft}(s, a; \boldsymbol{w})}\right)\right]^2.
  \label{eq:sacqloss}
\end{equation}
In these expressions, the Lagrange multiplier $\alpha$ is known as the temperature and determines the relative weights of the two terms of the new objective function.
It is an additional parameter that is iteratively adjusted at each step of the learning algorithm\footnote{The gradient descent step that follows equation \ref{eq:sacentropyloss} is run first, followed by the two steps optimising equation \ref{eq:sacpiloss} and equation \ref{eq:sacqloss}, so that the parameter update steps take into account how far the current strategy profile is from the uniform distribution and give proper weight to the regularisation term in objective functions \ref{eq:sacqloss} and \ref{eq:sacpiloss}.} by a constant proportional to
\begin{equation}
\underbrace{\mE_{\sigma(\boldsymbol{\theta})}\left[- \log \sigma(a|s; \boldsymbol{\theta})\right]}_{h(\sigma(a|s; \boldsymbol{\theta}))} - \bar{h}
  \label{eq:sacentropyloss}
\end{equation}
in order to keep the expected KL divergence between the estimated strategy profile and a uniform distribution close to $\bar{h}$.
This rule increases the temperature $\alpha$ when the objective is negative, i.e.  $\mE_{\sigma(\boldsymbol{\theta})}[- \log \sigma(a|s; \boldsymbol{\theta})]$ is smaller than the target KL divergence $\bar{h}$, and decreases it when it is positive.

\subsection{Economic environment}

I follow the economic environment of \citet{calvano_artificial_2020} and evaluate the performance of soft actor-critic on repeated Bertrand competition.
The action space is defined by the prices set by the $n$ agents, the state space is defined by the actions of all agents in the last period, and the reward is given by profits under logit demand and constant marginal costs.

In a given time step $t$, the price set by agent $i$ is a positive real number $p_{i, t}$.
Each of the $n$ firms produces a distinct product $i$.
Demand for product $i$ is given by
\begin{equation}
  q_{i, t} = \frac{e^{\frac{a_i - p_{i, t}}{\mu}}}{\sum_{j=1}^n e^\frac{a_j - p_{j, t}}{\mu} + e^{\frac{a_0}{\mu}}}.
\end{equation}
Parameters $a_i$ and $\mu$ capture vertical and horizontal differentiation.
The profit/reward function is
\begin{equation}
  \pi_{i, t} = (p_{i,t} - c_i) q_{i, t}
\end{equation}
where $c_i$ is marginal cost.

To ensure that the dimension of the state space is finite and does not increase with time, I posit a bounded memory of agents' past prices.
I define the state space as the set of histories of length $k$, obtaining a state space of dimension $nk$.
Unless otherwise specified, I focus on the case where $k = 1$, that is, agents only remember the last price that they and their competitors played.
As soon as memory is bounded, the process becomes nonstationary, thus invalidating known proofs of convergence based on the policy gradient theorem \citep{sutton_reinforcement_2018}.
However, convergence is not explicitly ruled out and can still be verified experimentally.

As mentioned earlier, I do not discretise the model, because policy gradient methods can operate in continuous state and action spaces.
To ensure prices are above marginal cost, I scale the actor network output $x_t \in (-1, 1)$ as $p_t = \frac{x_t + 1}{2} \cdot (\overline{p} - \underline{p}) + \underline{p}$, where $\overline{p}, \underline{p}$ are the maximum and minimum prices.
I denote the Nash and monopoly prices as $p_N, p_M$ and set $\underline{p} = p_N - \xi(p_M - p_N), \overline{p} = p_M + \xi(p_M - p_N)$, with $\xi \in \mR_+$ a parameter.
This means that the action space is $\mathcal{A} = (\underline{p}, \overline{p})$ and the state space is $\mathcal{S} = \mathcal{A}^n$, with $n$ being the number of agents.

To characterise emerging collusion under autonomous pricing as ``tacit'', I want to ensure that there is no possibility of communication between the two agents, or any other information that could potentially be described as nudging the algorithms towards collusive behaviour.
For this reason, I do not assume complete information and let agents learn in a completely unsupervised and model-free way.
I do not make the agents share any parameters, do not inject any knowledge about the environment, and only let agents observe their own profits and their and their competitors' market prices.
The only objective of the algorithms is to maximise profits.

\section{Results}
\label{chap:results}

In this section, I describe the evolution of payoffs obtained in the environment at hand by the algorithms specified in the previous section.
I also analyse what kind of strategy profiles they end up learning, whether they reach strategies that make collusive prices ``incentive compatible'', and the reasons for their emergence.

\QuoteRtwoFailureLearning

From the perspective of repeated games, the folk theorem implies multiple sustainable long-run equilibria, including competitive and collusive outcomes.
Algorithmic pricing can therefore be viewed as a way to select one among such equilibria in a non-supervised way.
However, it is not an equilibrium for all agents to always post high prices because such prices are not supported by any punishment and such a strategy could be easily exploited by an opposing firm by undercutting the agent when it unconditionally prices high.
The aim of this section is therefore to separate unconditionally high pricing from learned reward-punishment behaviour by testing whether a one-period static-best-response deviation is profitable and by visualising policy responses to other one-period deviations across the price space.
Even if the main focus of the paper is the speed of convergence, responses to deviations remain relevant to assess whether fast convergence comes at the expense of failure to optimise.

Reported baseline results have been obtained using the hyperparameters specified in table \ref{table:hyppar}.
Supra-competitive profit gains and reward-punishment behaviour are still present after changing these hyperparameters, even if changes are larger than one order of magnitude.
Varying the hyperparameters or using different function-approximation algorithms\footnote{In particular, for example, proximal policy optimisation \citep{schulman_trust_2015} reaches profit gains higher than 80\% in around 5,000 periods, but it does so by unconditionally high pricing instead of reward-punishment schemes -- this is expected, as on-policy algorithms tend to perform worse in terms of generalisation.
Given the recent progress in on-policy reinforcement learning, driven by their use in fine-tuning large language models, and their relative absence from the algorithmic pricing literature -- Q-learning, deep Q-learning, and soft actor-critic are all off-policy -- studying their behaviour in algorithmic pricing settings is an interesting avenue for future work.} primarily affects the fraction of sessions verifying incentive compatibility.
In terms of economic characteristics, to enable comparisons with the Q-learning of \citet{calvano_artificial_2020}, I fix inverse aggregate demand $a_0 = 0$, horizontal differentiation  $\mu = 0.25$, quality indices $a_1 = a_2 = 2$, and price space margin  $\xi = 0.1$.

As in \citet{calvano_artificial_2020}, the definition of ``convergence'' in reinforcement learning is not trivial.
Usually, articles in the reinforcement learning literature deem convergence as achieved if reward exceeds some predefined threshold for a certain number of time steps, usually an index of how well humans do in a given task.
This is not appropriate for algorithmic pricing, as doing so would potentially amount to \emph{requiring} high prices from the algorithms.
In addition, given that prices are continuous and agents' strategy profiles are required to have non-zero variance by the constraint, it is not trivial to define when they have ``stabilised'': soft actor-critic, by definition, converges to a stochastic strategy profile that does not maximise average profit only but also constrains its variance, therefore it can never converge to a deterministic strategy profile, or to one with arbitrarily low variance.

\QuoteRtwoConvergenceCriterion

\QuoteRtwoDeterministicEvaluation
However, about half of the sessions enter a cycle after a deviation from the steady state.
All of these cycles are of period 2, mirroring the cyclical pattern reported in \citet{calvano_artificial_2020}.
In all cases, the amplitude of the cycle is low, and it dampens as the system returns to its fixed point, as detailed below.

\subsection{Training curves}
\label{sec:training-curves}

In this section, I show algorithms' behaviour during learning.
I show that the pricing pattern is similar to \citet{calvano_artificial_2020}, with algorithms first cutting then raising prices.

\begin{figure}[t]
  \caption{Profit Gain Over Time}
  \label{fig:pgcurve}
  \centering
  \includegraphics[width=\textwidth]{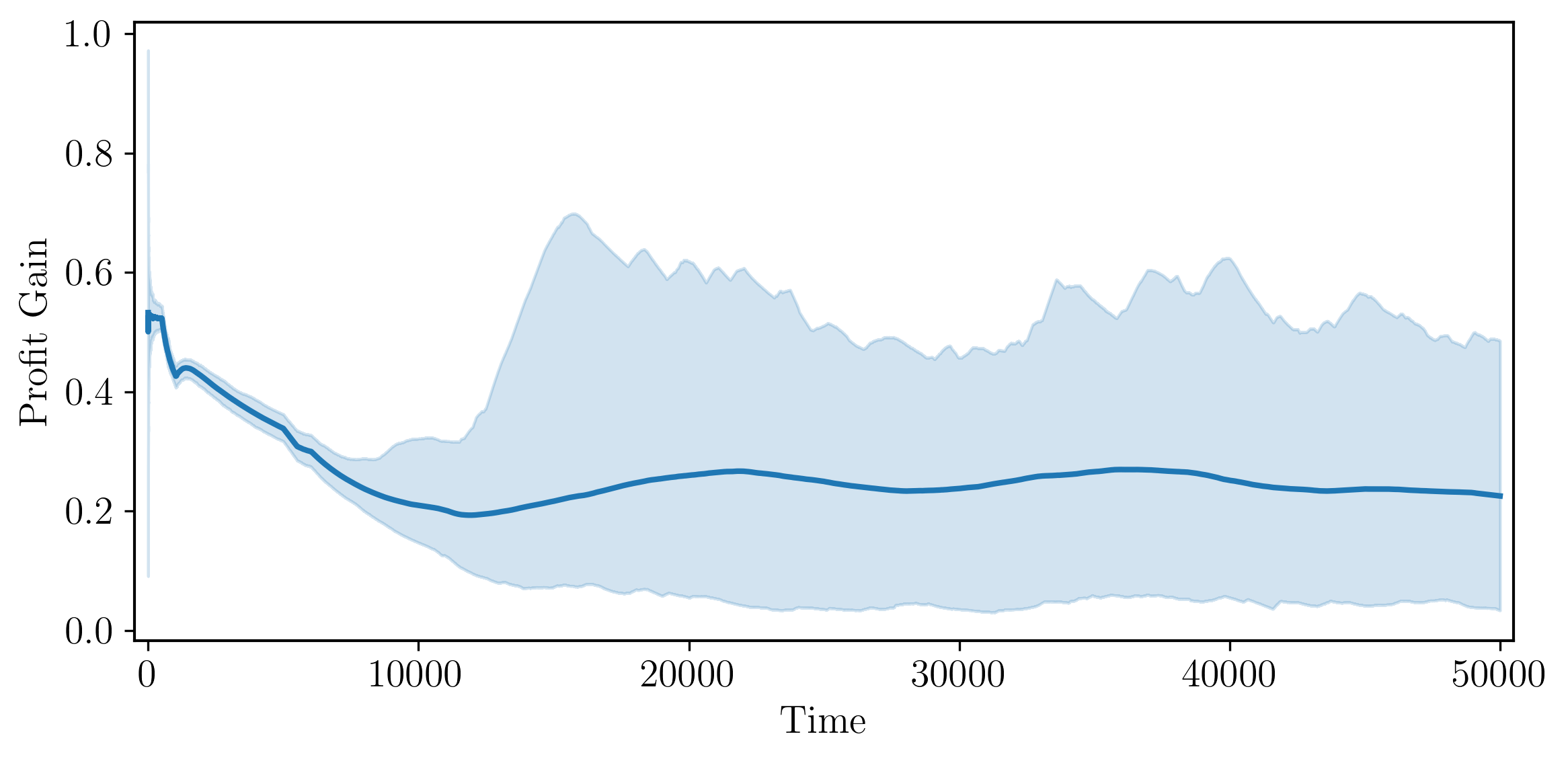}
\floatfoot{\textbf{Notes:} Lines show 1{,}000-step moving averages of the average profit gain $\Delta = \frac{\bar{\pi} - \pi_N}{\pi_M - \pi_N}$, where $\bar{\pi}$ is profit, defined relative to the Bertrand-Nash profit $\pi_N$ and monopoly profit $\pi_M$, across 100 random seeds; the shaded band shows the empirical 95\% confidence interval: since there are 100 seeds, for each timestep $t$, the lower limit is the mean between the 2nd and 3rd prices at $t$ and the upper limit is the mean between the 97th and 98th prices at $t$.}

\end{figure}

I employ the same normalised measure of average profit gain described in \citet{calvano_artificial_2020}:
\begin{equation}
  \Delta = \frac{\bar{\pi} - \pi_N}{\pi_M - \pi_N}
\end{equation}
where $\bar{\pi}$ is average profit over the relevant evaluation window, $\pi_N$ is the Bertrand-Nash static equilibrium profit, and $\pi_M$ is monopoly profit.
$\Delta$ is 1 in case of monopoly pricing and 0 in case of Bertrand-Nash play.
Profits themselves cannot be negative as I assume no fixed costs and because the lowest price agents can set is higher than marginal cost.
However, if profits are lower than Nash equilibrium profits, $\Delta$ will be negative.
I plot the training curve of an average of 100 experiments with different random seeds, showing profit gains over time, in Figure \ref{fig:pgcurve}.

Initially, agents appear to be engaging in competitive pricing, driving profits down closer to the Bertrand-Nash equilibrium level of 0, though they do not reach perfect competition.
This echoes the behaviour shown by Figure 10 in \citet{calvano_artificial_2020}, with prices falling closer to Bertrand-Nash before rising to higher levels.
The trend line then reverses its decline and begins a steady upward climb.
However, the lower bounds dip noticeably before recovering, and the wide spread suggests that while the average outcome has supra-competitive profits, the path to high prices is volatile.

\begin{figure}[t]

  \caption{Profit Gain Distribution}
  \label{fig:pg}
  \centering
  \includegraphics[width=\textwidth]{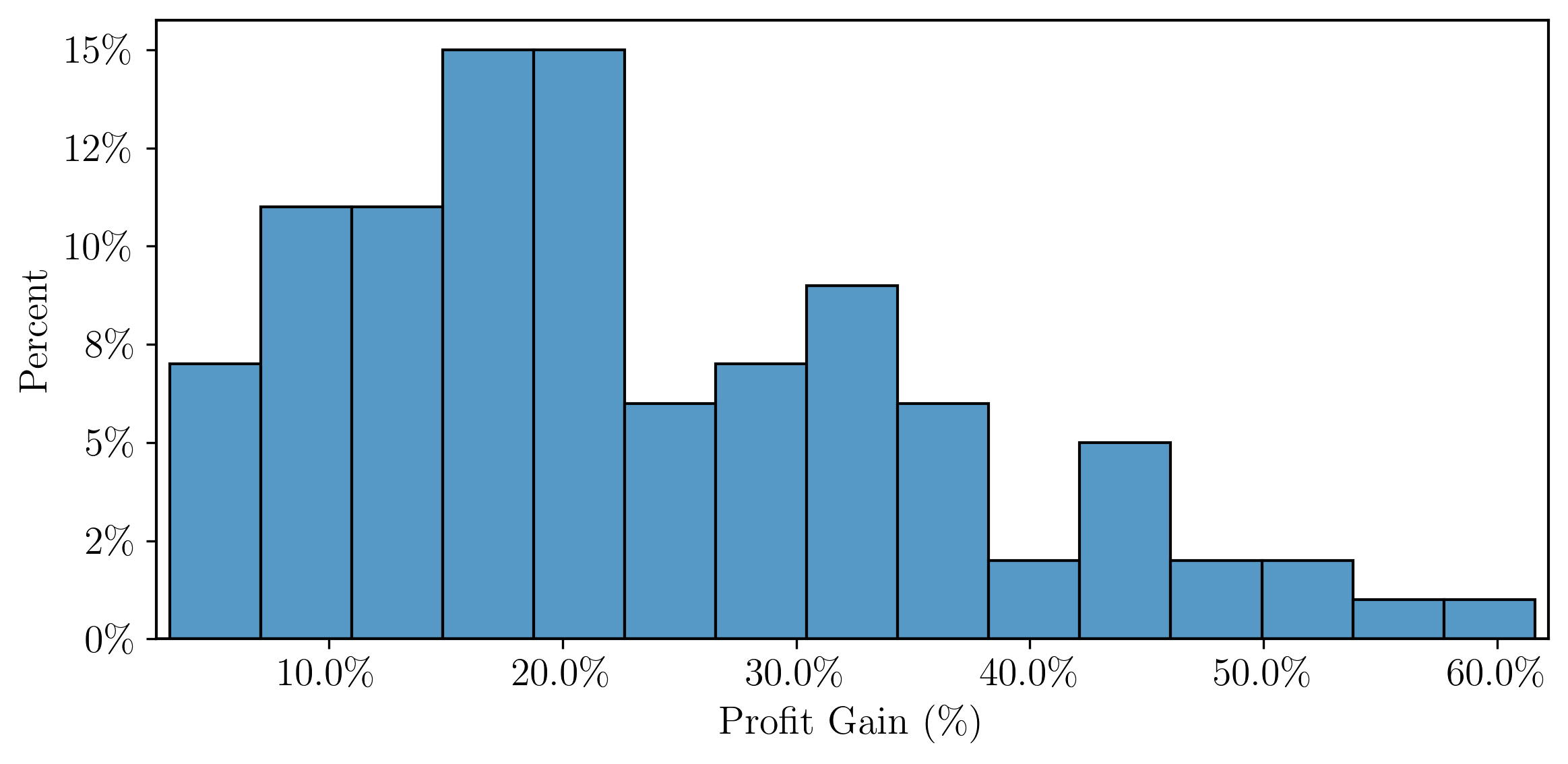}
  \floatfoot{\textbf{Notes:} Histogram across 100 runs of the profit gain per session, averaged across agents at the end of the simulation.
  Profit gain equals $(\bar{\pi}-\pi_N)/(\pi_M-\pi_N)$ so zero corresponds to Bertrand-Nash and one to monopoly profits.}
\end{figure}

Usually, deep reinforcement learning algorithms tend to exhibit notable instability and high variance, which is usually masked by averaging across random seeds but emerges when the variation is visualised.
To understand the degree of variation, we can look at the distribution of profit gains in Figure \ref{fig:pg}.
Profit gains range between 10\% and 60\%, with a mean of about 30\%.
Indeed, not every run achieves the same high level of coordination, and the histogram shows that the distribution of profit gains has a high variance and is right-skewed, with most runs achieving levels of profit gain comparable to the uniform pricing benchmark of 40\%, and about one in ten sessions reaching gains higher than 50\%.
That said, it is important to evaluate whether these profit gains are sustained by off-path incentives, or just the result of chaotic pricing or unconditionally high pricing.
The next section will show that sessions are about evenly split between the two.

\subsection{Equilibrium play}

Having verified behaviour at the end of the simulation, one key question in reinforcement learning is how learned strategy profiles generalise to unseen behaviours.
This same question arises in economics, where a collusive strategy is sustained only if it is not optimal to deviate from it, i.e. if the non-deviating agent acts in such a way as to punish deviations and render them unprofitable.
Indeed, as \citet{den_artificial_2022} shows, it is not an equilibrium to use algorithms in the first place if deviating from the learned pricing strategy is optimal in the long run, and supra-competitive prices may not be determined by actual collusive strategy but by failure to learn an optimal strategy profile and mere unconditionally high pricing.
To understand whether agents learned optimal behaviour even off the path of equilibrium, in this section I analyse agents' responses to deviations.

During learning, each algorithm $i$ learns to map prices $(p_{1, t}, p_{2, t})$ at time $t$ to a mean  $\mu_{i, t+1}$ and a standard deviation $\Sigma_{i, t+1}$.
Then, prices in $t+1$ are sampled from $\mathcal{N}(\mu_{i, t+1}, \Sigma^2_{i, t+1})$.
To evaluate responses to a deviation, I disable learning, so that each agent's function $(p_{1, t}, p_{2, t}) \mapsto (\mu_{i, t+1}, \Sigma_{i, t+1})$ does not change, and let agents play their mean action $\mu_{i, t}$ for 50 periods.
\QuoteRtwoSettlingWindow
Then, I force one of the two agents to defect, for one period,  to the static best-response to the last price played by the rival.
I denote by $t = 0$ the time of defection, so that if agent $i$ is the defector and agent $-i$ is compliant, $p_{i, t}$ is the one-period best response to $p_{-i, t-1}$.
After the deviation, I let agents play their mean action $\mu_{i,t}$ again.

\begin{figure}[t]
  \caption{Discounted Deviation Payoffs}
  \label{fig:discdev}
  \centering
  \includegraphics[width=\textwidth]{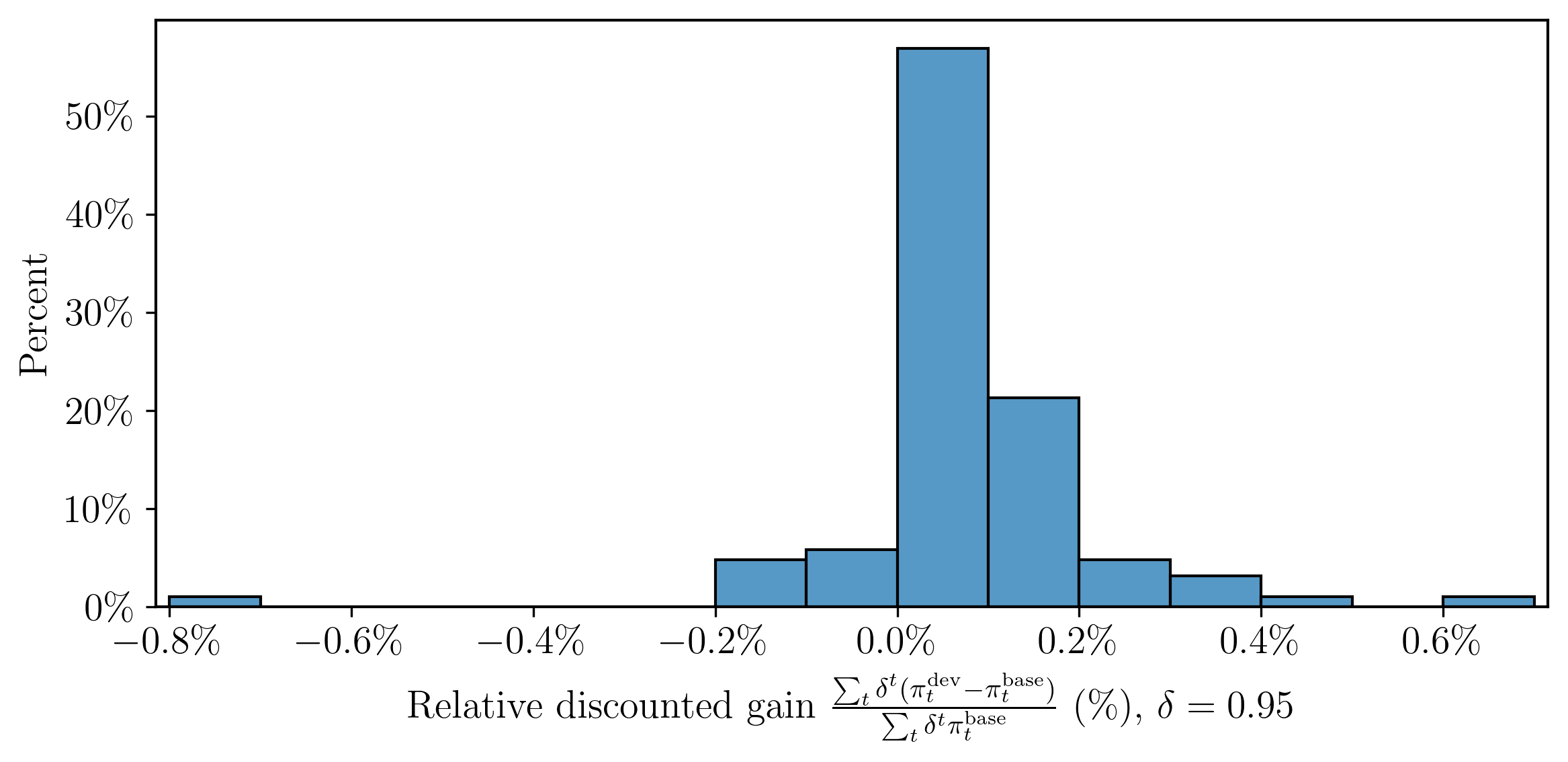}
  \floatfoot{\textbf{Notes:} The x-axis variable, relative discounted gain, quantifies the percentage change in profitability over the 100-period post-deviation evaluation path, including the deviation period itself.
Profits after deviation are discounted at $\delta=0.95$ to match \citet{calvano_artificial_2020}.
  It compares the average profit obtained after deviation to the profit obtained in the counterfactual with no deviation.
  Deviation profits are measured after one agent is forced to defect to a static best-response strategy against the rival's last played price.
  Bars are histogram bins, so the central bar corresponds to deviations with gains close to zero rather than deviations with exactly zero gain.
  A value near zero suggests that the deviation to a static best-response yielded no advantage or disadvantage compared to the learned strategy.
  Values deviating from zero (the tails) represent scenarios where the static best-response defection successfully exploited the rival (positive gain) or inadvertently triggered a punishment or suboptimal outcome (negative gain) relative to counterfactual, deviation-free play.}
\end{figure}

We can then look at the distribution of discounted\footnote{The discount factor is extraneous to the learning process in the average-reward formulation, but it would not be meaningful to look at the difference in differential profits, because the lack of discounting means that the time horizon at which we evaluate them matters.
To fix ideas, suppose that the defector gains 20\% from a one-period deviation, that it is not punished, and that they both return to cooperation forever.
Then the differential deviation gain $\frac{1}{T} \sum_t (\pi^{dev}_t/\pi^{base}_t - 1)$ after $T$ periods is 10\% if $T = 2$, 0.2\% if $T = 100$, and 0.02\% if $T = 1000$.
Conversely, with an evaluation discount factor $\delta = 0.95$, the discounted gain is 10.26\% if $T = 2$, and 1\% for all $T > 100$.}
profits originating from a deviation, shown by the histogram in Figure \ref{fig:discdev}.

Looking at discounted profits is particularly interesting because, as far as the agent is concerned, future payoffs are \emph{not} discounted in the average reward formulation.
\QuoteRtwoGamma
As shown by Figure \ref{fig:discdev}, the distribution of deviation gains exhibits a large mass near 0 profits, and a few outliers are enough to make the mean a very unreliable statistic.
\QuoteRtwoHistogram

However, this is an entirely exogenous choice and mechanically more deviations become profitable when the evaluation discount factor falls because learning does not take into account the discount factor as it does in \citet{calvano_artificial_2020}.
In the limit where $\delta = 0$, any deviation is regarded as profitable, so no collusion is possible; when instead $\delta = 1$, discounted profits are analogous to differential profits because the quantity $\pi(S_t, A_t) - \bar{\pi}(\sigma)$ becomes 0 as $t \rightarrow \infty$ if policy estimation converged and policies are evaluated deterministically.
As \citet{naik_discounted_2019} show, it is not a well-defined optimisation problem to maximise discounted payoffs under function approximation.
For this reason, evaluating average-reward reinforcement learning agents based on how they minimise discounted loss from a deviation is not entirely fair, as they were not optimised for this objective.

Figure \ref{fig:discdev} shows that even in the common case in which it is profitable to deviate, it only leads to at most a 1\% gain relative to the counterfactual with no deviation.
Most sessions cluster near zero gains, but a modest right tail is enough to keep the share of profitable deviations above one half for most of the training path.
Table \ref{table:devlearn} reports the distribution of differential deviation gains at multiple training checkpoints and the share of tested agent-level deviations that are unprofitable.
I classify a tested agent-level deviation as ``incentive compatible'' when this static-best-response deviation is not profitable.
The table indicates that roughly 41\% of tested agent-level deviations are unprofitable by this metric at 10,000 steps, when prices are still close to Bertrand-Nash.
The unprofitable share is about 29\% at 30,000 and rises to about 36\% by 50,000 when some agents learned to maintain supra-competitive prices with off-path punishments.
Longer simulations show the same pattern, and cases that have learned a reward-punishment scheme after 50,000 periods tend not to forget it until the replay buffer is completely filled with identical experiences.
In what follows, I thus focus my attention on the cases that pass this one-period deviation test.
This restriction overstates the reliability of collusive learning if read as a statement about all runs, since firms are less likely to deploy algorithms that are known to fail to settle; I return to this in Section \ref{chap:discussion}, and in Section \ref{chap:conclusions} I point to future research directions, such as rematching and sequential pricing, that could reconcile these concerns.
At the same time, these simulations provide suggestive evidence that it is possible to gauge convergence empirically for both firms and antitrust authorities, because equilibrium play is positively correlated with profit gains (Pearson's coefficient 0.20, compared to 0.12 in \citealt{calvano_artificial_2020}), that is, when a session converges to equilibrium play, it tends to converge to higher prices.

\begin{table}[t]
  \caption{Deviation Gains During Learning}
  \label{table:devlearn}
  \centering
  \resizebox{\textwidth}{!}{
  \begin{tabular}{lccccc}
\hline
Step & 25th percentile (\%) & Median (\%) & 75th percentile (\%) & Mean (\%) & Unprofitable \% [95\% CI] \\
\hline
\num{10000} & -0.0042\% & 0.0012\% & 0.0083\% & 0.1\% & 41\% [35\%--49\%] \\
\num{20000} & -0.0011\% & 0.0042\% & 0.035\% & 0.11\% & 32\% [26\%--39\%] \\
\num{30000} & -0.00028\% & 0.0071\% & 0.044\% & 0.22\% & 29\% [23\%--36\%] \\
\num{40000} & -0.0013\% & 0.0059\% & 0.033\% & 0.085\% & 31\% [24\%--38\%] \\
\num{50000} & -0.0017\% & 0.0029\% & 0.017\% & 0.06\% & 36\% [30\%--43\%] \\
\hline
\end{tabular}

  }
  \floatfoot{\textbf{Notes:} Each row summarizes the distribution of differential deviation gains at the indicated training step, computed using the same one-period static best-response deviation as in Figure \ref{fig:discdev}.
  For each training session and each agent, I first take the last 50 realized prices prior to the checkpoint and use their average as the state input; in a separate evaluation session that does not impact training, the agents then play for 50 steps to settle into their learned strategies.
  One agent deviates to the static best-response price against the rival's last played price, while the rival follows its learned policy; from then onward both follow their learned policies.
  The differential deviation gain for the deviating agent is $\frac{1}{100}\sum_{t=0}^{99}\left(\frac{\pi^{\text{dev}}_{t}}{\pi^{\text{base}}_{t}}-1\right)$, where $\pi^{\text{dev}}_{t}$ is the deviator's profit in the deviation path and $\pi^{\text{base}}_{t}$ is the profit in the counterfactual path with no deviation.
  The percentiles and mean are computed across all agent-level gains pooled across sessions at that step.
  ``Unprofitable \%'' is the share of these gains that are non-positive.
  The 95\% confidence interval is obtained by bootstrapping the pooled gains with replacement for 5{,}000 iterations, recomputing the unprofitable share each time, and reporting the 2.5th and 97.5th percentiles of that bootstrap distribution.}
\end{table}

\subsection{Responses to deviation}

\begin{figure}[t]
  \caption{Emergent Reward-Punishment Pattern}
  \label{fig:ir}
  \centering
  \includegraphics[width=\textwidth]{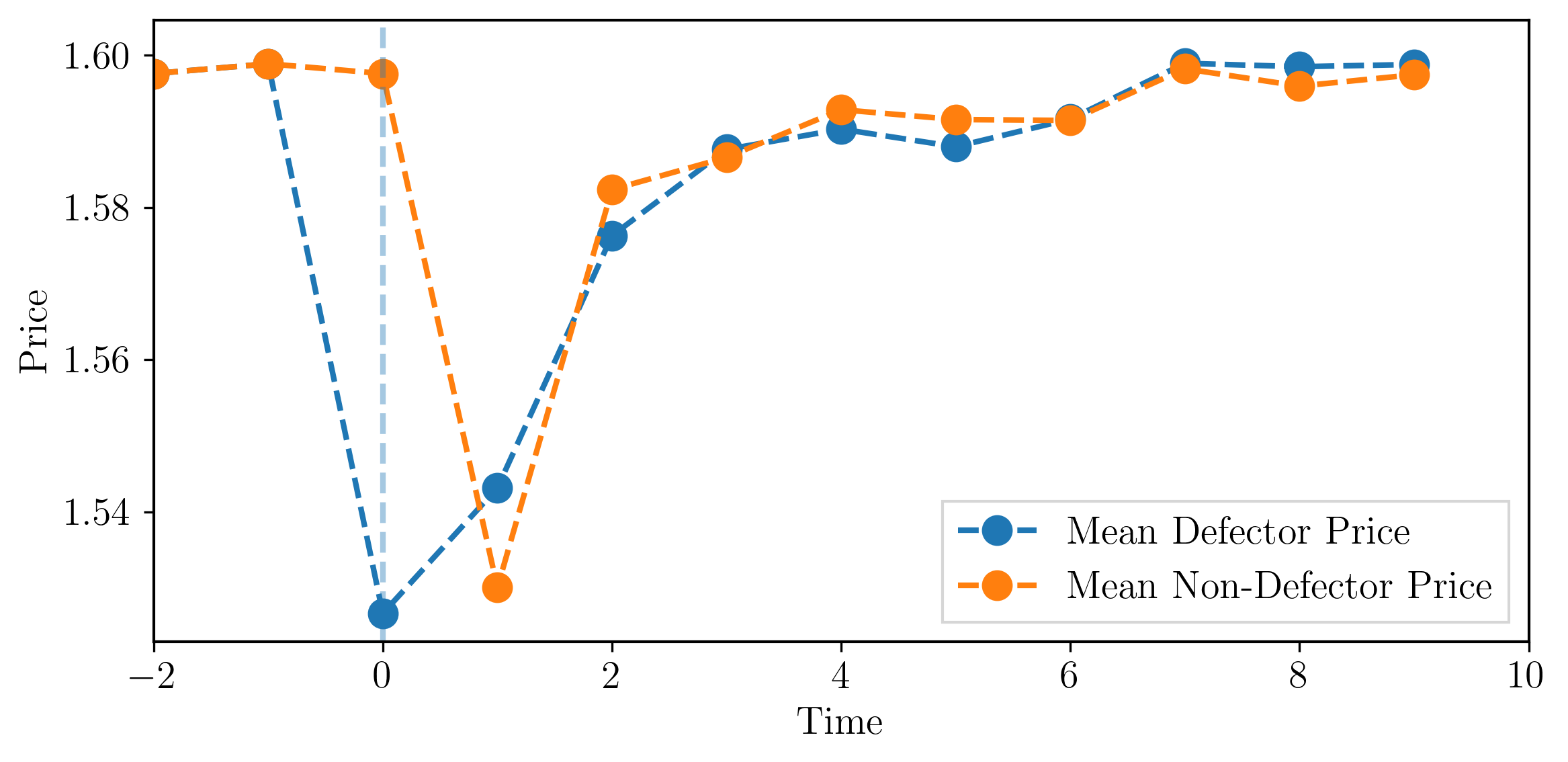}
\floatfoot{\textbf{Notes:} Average price paths when one agent deviates for one period to the static best response; agents play their mean strategy for 50 steps before one of them is forced to deviate. The plot includes only sessions in which the one-period static-best-response deviation is not profitable. As every agent in each seed has a different strategy profile, this plot shows the average of played prices at each $t$. }
\end{figure}

To understand \emph{how} equilibrium play is maintained in the sessions where it is reached, I let each agent defect in each of these sessions and plot the average price played after the deviation in each time step across agents and sessions in Figure \ref{fig:ir}.
The agents are not learning a grim trigger strategy with permanent punishment, but rather a one-period punishment followed by gradual return to higher prices.
Given that the agents have a bounded memory, they are unable to keep track of the status of the punishment directly.
Instead, reversion to cooperation is achieved by learning to match each other's price and to repeatedly respond to a matched sub-competitive pricing with another matched pricing strategy with slightly higher prices, until the pre-deviation prices are reached again.
As we shall see, this only happens if the raise in price is matched by the two agents.

\subsection{Reward-punishment across random seeds}
\label{sec:performance-seeds}

In this section, I show that the reward-punishment scheme is not an artifact of averaging but happens in all simulations where equilibrium play is attained\footnote{About 20\% of simulations with profitable deviations still show reward-punishment schemes, but these are not strong enough to deter the tested deviation and they do not get stronger with more learning. I take a conservative stance and exclude those simulations.}.

\begin{figure}[t]
\caption{Impulse Response Boxplot}
  \label{fig:boxir}
\centering
    \includegraphics[width=\textwidth]{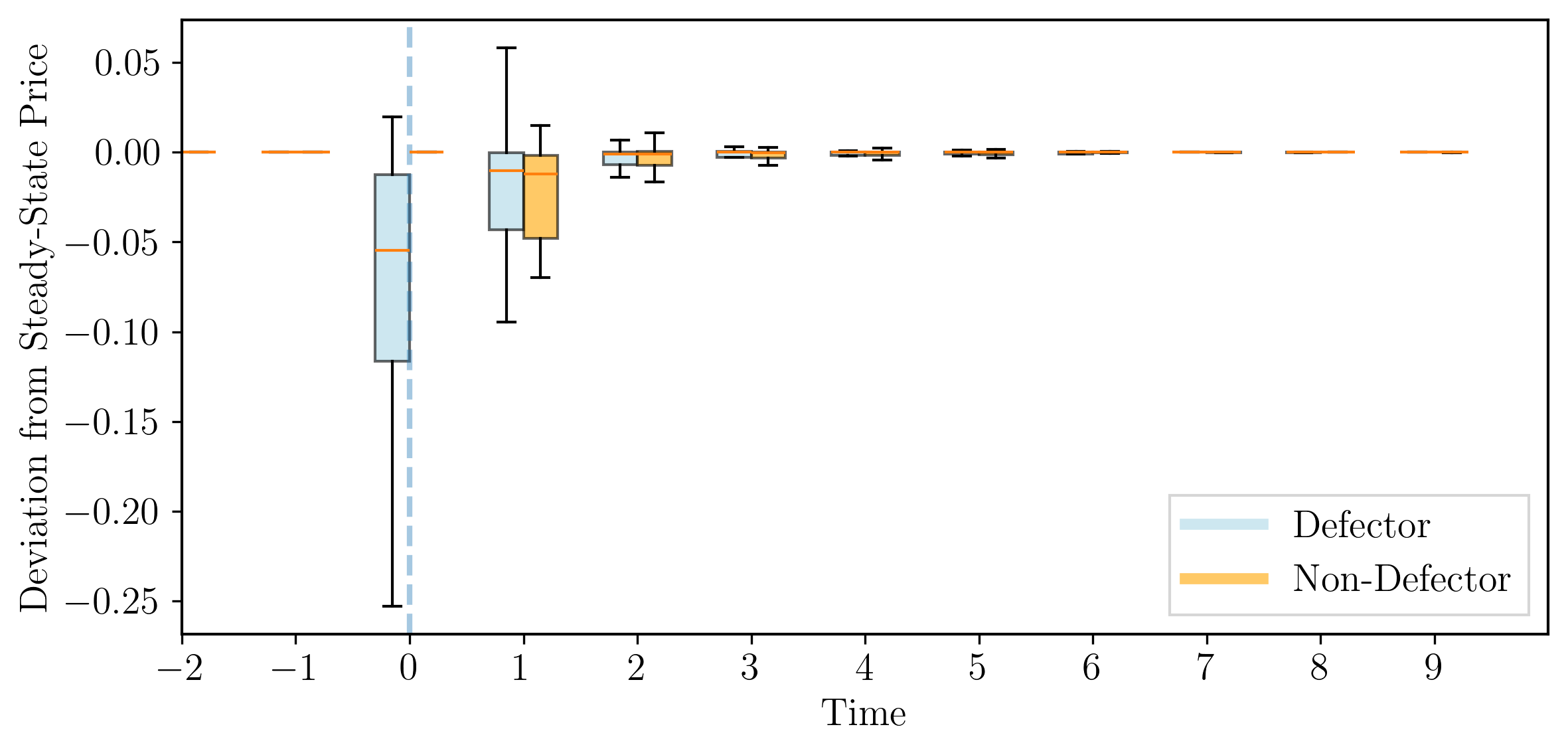}
\floatfoot{\textbf{Notes:} Box plots summarise the distribution, across sessions passing the one-period deviation test, of the change in prices relative to the pre-deviation price for the compliant agent (right) and the deviator (left) after a one-period static-best-response deviation; whiskers cover the 5th--95th percentiles.}
\end{figure}

To discern whether the averaging of impulse responses is masking more complex behaviour, we can inspect the distribution of the difference between the price played just before the deviation and the price played in the subsequent time steps, shown in Figure \ref{fig:boxir}.
These box plots confirm that the reward-punishment scheme is actually emerging in all retained runs and is not just a result of averaging chaotic impulse responses, although what varies between runs, as we shall see later, is how harsh this learned punishment is.
They also show that reward and punishment are much more stable and with lower variance than what is obtained by Q-learning in \citet{calvano_artificial_2020} and shown in their Figure 5.
Prices rarely surpass the starting equilibrium level and exhibit much lower variance overall; when they do surpass the equilibrium level, the difference is slim.

\subsection{Visualising learned strategy profiles}
\label{sec:learned-policies}

The static best-response is not the only deviation possible.
In principle, agents could deviate for one period to any other price, and a collusive strategy should respond in a way that makes such deviations unattractive.
Some examples of deviations are: deviation to monopoly price, an upwards deviation that disrupts equilibrium and should therefore be punished, however ``advantageous'' for the other agent; deviation to marginal cost, similar to the optimal punishment in \citet{abreu_extremal_1986}; deviation to the Nash equilibrium price, the harshest reasonable punishment in a Bertrand setting studied in the previous section.
In this section, I thus focus on a single representative seed to show the behaviour of the policy throughout the price space, and show that agents return to one of two fixed points after one-period deviations from the plotted states.

Neural networks can learn any strategy profile and action-value for which functional forms\footnote{Under some assumptions on the neural network's structure, such functional forms may be discontinuous, as shown by \citet{ismailov_three_2023}.} exist \citep{hornik_multilayer_1989}.
Using continuous pricing in a game with two agents and one-period memory allows for visualising this functional form through a phase diagram, with observed prices $p_1, p_2$ as the axes, and the system dynamics represented by paths starting from various initial conditions.
This is possible because, in a duopoly, the mean of a strategy profile is effectively a function from both agents' prices in $\mathbb{R}^2$ to each agent's future price in $\mathbb{R}$ and, when turning off exploration by setting the variance to zero, strategies become deterministic; consequently, the phase diagram depicts the exact evolution, or orbits, of future prices based on current observations.
A similar visualisation is provided for Q-learning by \citet{calvano_artificial_2020} in their Figure 9.
\QuotePhaseDiagramMethod

\begin{figure}[t]
  \caption{Phase Diagram of a Learned Strategy Profile}
  \label{fig:hm}
\centering

  \hspace{-0.5cm}\includegraphics[width=0.7\textwidth]{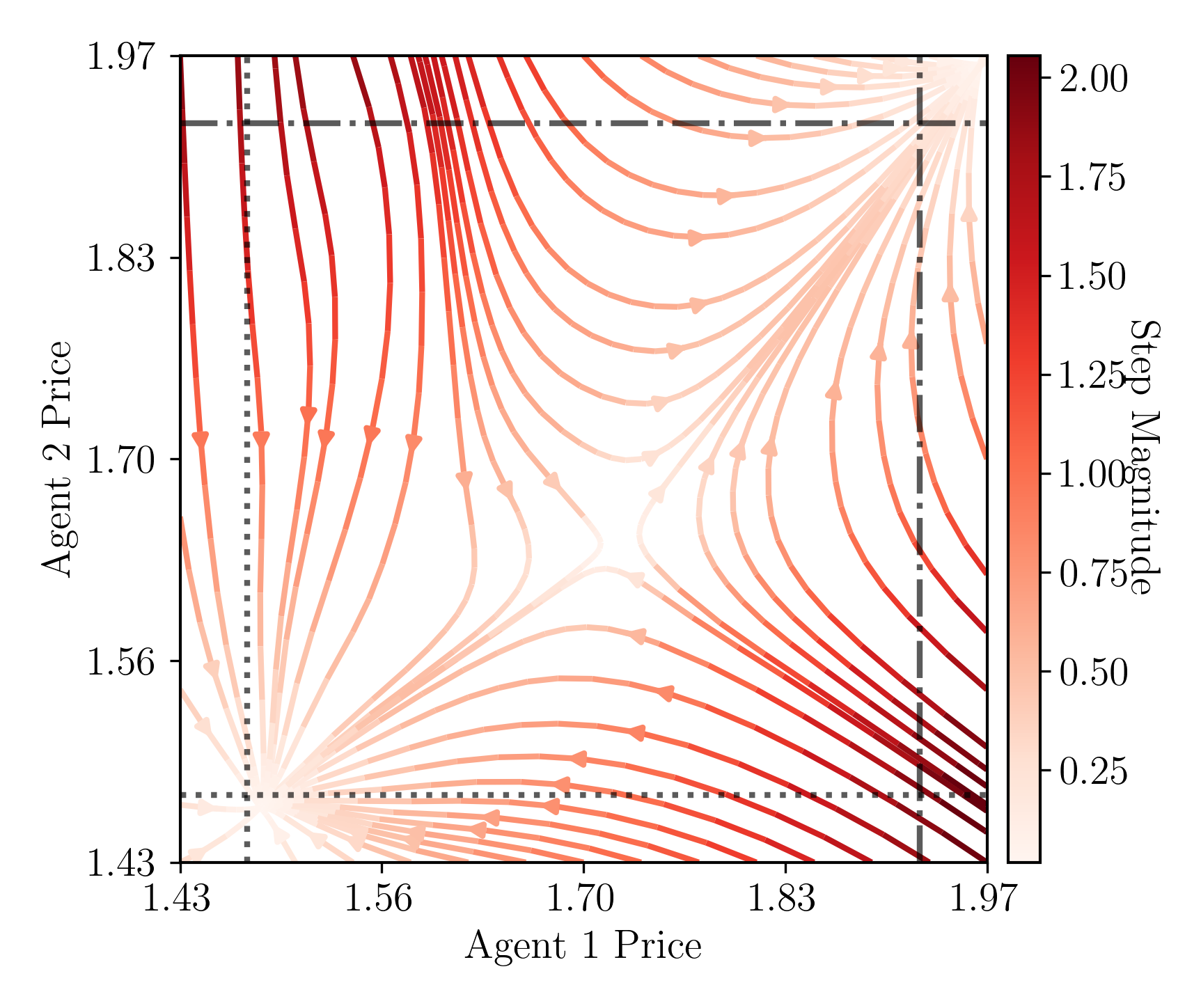}
  \floatfoot{\textbf{Notes:}
    The phase diagram visualizes the joint evolution of prices for Agent 1 ($x$-axis) and Agent 2 ($y$-axis) based on a snapshot of the strategy profile at the end of the simulation, for a single seed.
    Other seeds are qualitatively similar but differ in the location of the fixed point(s), of which they have either one or two.
    Lines indicate the trajectory of future prices starting from any given state pair $(p_1, p_2)$.
    The color gradient denotes the step magnitude, or velocity, of the update vector, with lighter colors indicating regions where the strategy profile dictates smaller price adjustments.
  The dotted and dash-dotted lines represent the theoretical Bertrand-Nash and cooperation price benchmarks, respectively.
  \QuoteRtwoFixedPoints}
\end{figure}

Figure \ref{fig:hm} shows a phase diagram corresponding to the strategy profiles learned by two agents, for a representative experiment\footnote{It is not appropriate to average here as no agent actually follows the ``mean strategy profile''. For example, from the same state $(x, y)$, one seed will move to $(x', y')$, another seed to $(x'', y'')$, but no seed is guaranteed to move to the mean $\frac{1}{100} [(x', y') + (x'', y'') + ...]$.}.

\QuotePolicyResponse
There is also a second fixed point, closer to the Bertrand-Nash equilibrium.
This shows how, in this run, larger one-period deviations are punished with convergence to a low-price fixed point close to the Bertrand-Nash price, while smaller deviations are forgiven and lead to a gradual return to the high-price fixed point.

\subsection{Number of agents}
\label{sec:number-agents}

The competition model at hand leads to the conclusion that markets with more firms are harder to collude in, because monopoly profits are lower and therefore it is more tempting to deviate as punishment is weaker.
\QuoteRtwoThreeAgents
This result is consistent with what is obtained by Q-learning but requires no adjustments to the exploration strategy.

\begin{figure}[t]
  \caption{Three-Agent Profit Gains}
  \label{fig:pg3ag}
  \centering
  \includegraphics[width=\textwidth]{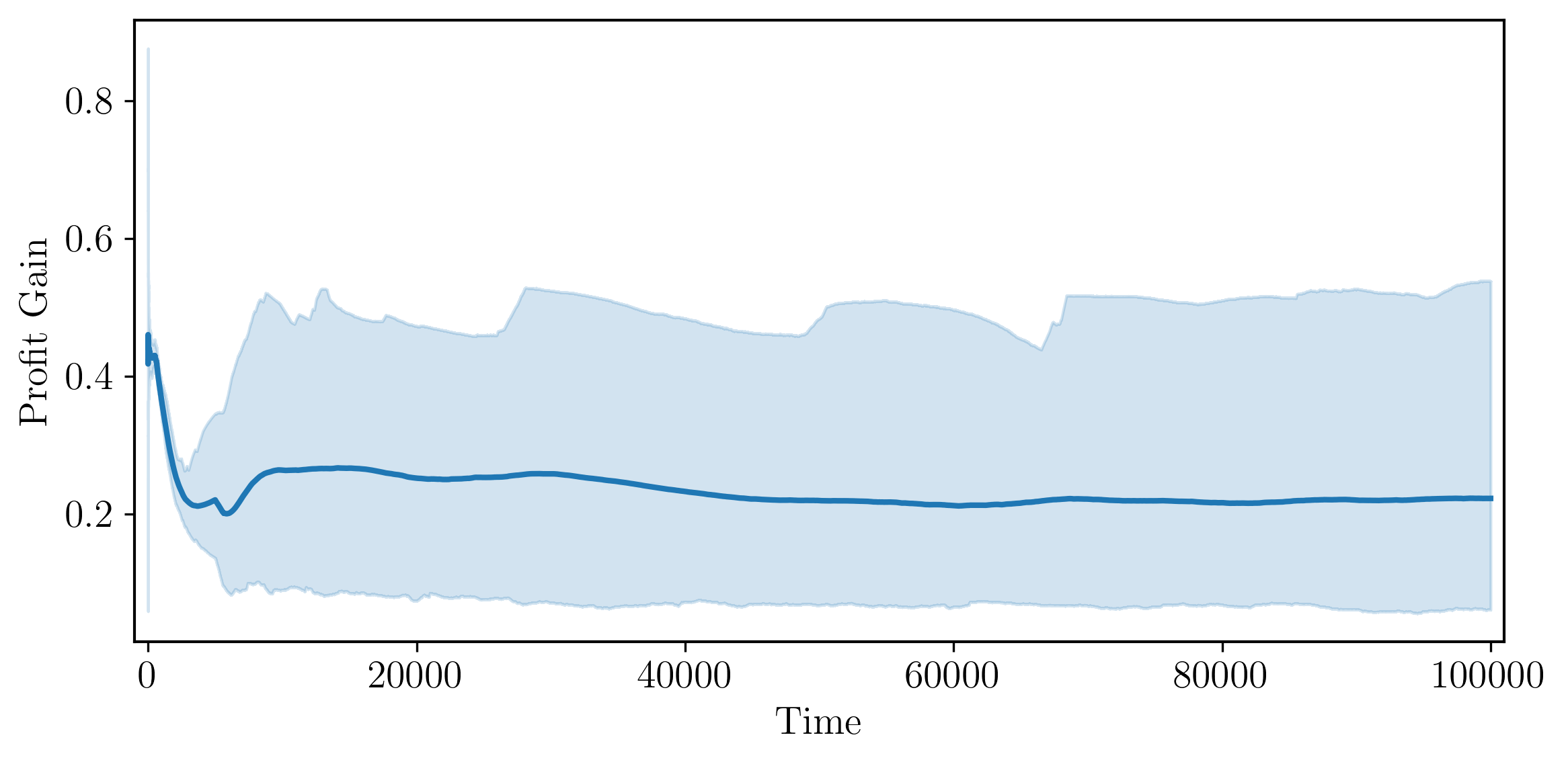}
\floatfoot{\textbf{Notes:} Lines show 1{,}000-step moving averages of the average profit gain $\Delta = \frac{\bar{\pi} - \pi_N}{\pi_M - \pi_N}$, where $\bar{\pi}$ is profit, defined relative to the Bertrand-Nash profit $\pi_N$ and monopoly profit $\pi_M$, across 100 random seeds in the three-agent case; $\pi_N$ and $\pi_M$ are recomputed for the three-agent case. The curve includes the initial exploration and low-price learning phase and should therefore not be read as an average payoff from deploying the final policy. The shaded band shows the empirical 95\% confidence interval: since there are 100 seeds, for each timestep $t$, the lower limit is the mean between the 2nd and 3rd prices at $t$ and the upper limit is the mean between the 97th and 98th prices at $t$.}
\end{figure}

\begin{figure}[t]
  \caption{Three-Agent Impulse Response}
  \label{fig:boxir3ag}
  \centering
  \includegraphics[width=\textwidth]{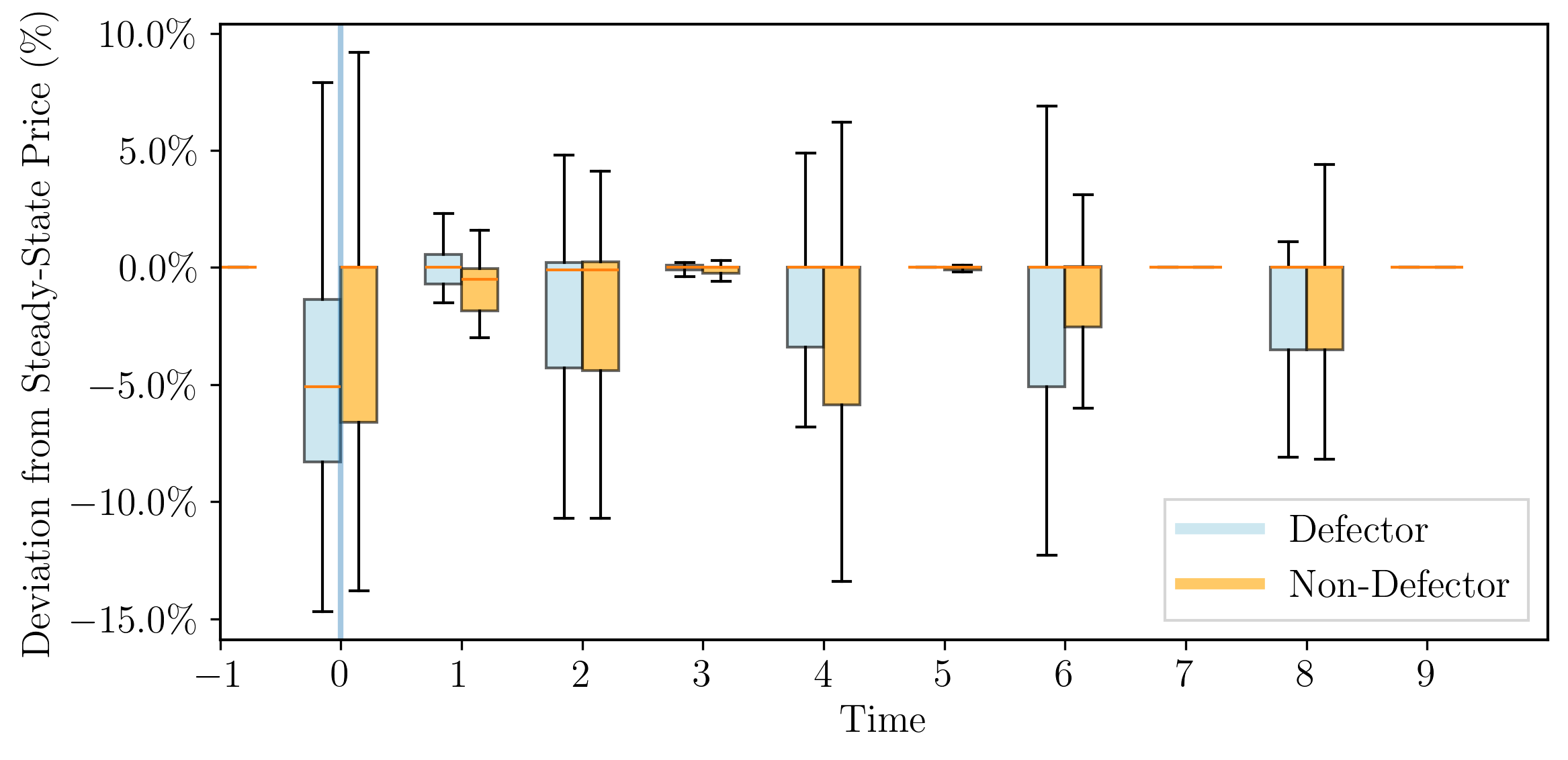}
  \floatfoot{\textbf{Notes:} Box plots summarise the distribution, across sessions passing the one-period deviation test, of the change in prices relative to the pre-deviation price for the deviator (left) and the non-defector (right) after a one-period static-best-response deviation in the three-agent case; the non-defector is the average of the non-defecting agents; whiskers cover the 5th--95th percentiles.}
\end{figure}

However, behaviour remains the same, with time to convergence being similar.
There is a higher variance in impulse responses, as shown by Figure \ref{fig:boxir3ag}.
Moreover, equilibrium play is only reached in 30\% of sessions, compared to 36\% with two agents.
With more than three agents, e.g. $n = 4$, profit gains are lower and learning is slower, with reward-punishment schemes emerging but taking a longer time, about 100,000 periods, to make a sizable chunk of deviations to the static best-response unprofitable.

\section{Discussion}
\label{chap:discussion}

To properly contextualise these results, it is helpful to view algorithmic pricing through the lens of a ``meta-game'' structure described by \citet{den_artificial_2022}.
The meta-game treats the selection of the pricing algorithm itself (e.g., Q-learning) as the strategic action, where firms evaluate and choose their software based on its expected performance against rival algorithms.
They argue that Q-learning algorithms may not be a stable equilibrium choice in this meta-game; if it is possible to exploit algorithms during training, rational firms would do so, deviating towards different pricing strategies.
This article focuses specifically on the subgame where firms have already committed to learning algorithms as an equilibrium choice, contributing to the solution of the meta-game by backward induction.
As \citet{den_artificial_2022} themselves have shown, this happens, for example, if the firm is patient, or the algorithms learn fast enough.

Within this subgame, previous critiques, such as those by \citet{asker_artificial_2022}, have argued that standard Q-learning is too slow to be a credible threat of collusion.
This article speaks directly to this critique.
Informing algorithms with economic structure as in \citet{asker_artificial_2022} makes them update the expected value of every feasible price in each period, not just the one actually chosen, by using estimated counterfactual profits.
This approach dramatically accelerates learning but typically causes prices to converge to the Bertrand-Nash equilibrium rather than the high, supra-competitive prices seen with asynchronous updating.
However, algorithms that are informed with economic structure pose less of a challenge at the policy level, as such algorithms could be clearly labelled as collusive and antitrust authorities can intervene to outlaw their use.
A key contribution of this article is demonstrating that such explicit modelling is not necessary for rapid convergence, as function-approximation approaches can help algorithms learn system dynamics in a non-supervised way, and achieve collusive outcomes on a commercially relevant timescale without the need for economic structure that may pose issues with antitrust policies.

\QuoteRtwoDiscussionCaveat

\QuoteRtwoMultiPeriod

\section{Conclusions}
\label{chap:conclusions}

In this article, I show that state-of-the-art reinforcement learning algorithms using artificial neural networks to estimate a strategy profile function autonomously learn collusive strategies in a repeated pricing game faster than Q-learning in a subset of runs.
Agents begin to raise prices as soon as they are allowed to, a behaviour that is not exhibited by Q-learning.
In successful runs, learned strategy profiles can sustain high prices through reward-punishment schemes that make it unprofitable to deviate from the collusive equilibrium.

From an algorithmic engineering perspective, it is interesting to note that reinforcement learning algorithms are, as of the time of writing, suboptimal choices for control problems, especially non-stationary ones, compared to model-based optimal control.
Research in the field is often based on flawed assumptions \citep{nota_is_2020,naik_discounted_2019} and problems that are studied in the literature routinely violate the Markov assumption, just like the pricing game.
Usually, algorithms that achieve sample-efficient control in other fields, such as robotics, are model-based and require complete information.
These algorithms converge in a handful of time steps, and some have also been applied to games of cooperation.
Some examples are \citet{crandall_cooperating_2018,crandall_towards_2014,moravcik_deepstack_2017,tassa_synthesis_2012,kuindersma_optimization-based_2015}.

Nevertheless, the fact that deep reinforcement learning is effective in some runs shows that price competition is a relatively simple control task and algorithms setting prices, even with limited information and without being exogenously swayed towards collusion, can converge to a cooperative solution.
This article shows that, compared with Q-learning, convergence can be achieved on a shorter time scale, although 50,000 hourly periods would still imply years of learning.
Moreover, policy gradient algorithms like soft actor-critic have been standard practice in reinforcement learning research for the last few years.
Therefore, from a competition policy standpoint, the results of this article serve the purpose of advancing the understanding of pricing algorithms beyond proofs of concept and up to speed with the state of the art in computer science.

Promising directions for future work include theoretical study of the properties of pricing algorithms based on deep reinforcement learning, in a similar fashion to \citet{banchio_adaptive_2022} for Q-learning and \citet{possnig_learning_2023} for actor-critic Q-learning with discrete states.

\QuoteSequentialExtension

Furthermore, it would be interesting to see whether fine-tuning an actor-critic algorithm that learned a strategy profile function in a given session is sufficient to achieve collusion when it competes with an agent that learns from scratch, to simulate what is known in the literature as ``offline learning'' \citep{asker_artificial_2022}.
Indeed, offline learning garnered attention because exploration is costly to the firm, and extensive exploration is required to reach collusion when using Q-learning.
If firms are able to reach collusive outcomes solely with online learning algorithms on shorter horizons than tabular Q-learning, this may call for a re-evaluation of the cost of exploration and thus of the motivation for offline learning.

Finally, while \citet{assad_algorithmic_2023} provide empirical evidence of algorithmic collusion, it remains an open question whether the randomised trial-and-error of RL accurately reflects these observed learning processes.
Future research could bridge this gap by simulating learning from scratch in environments with entry and exit dynamics and comparing it to observed pricing patterns.

\bibliographystyle{chicago}
\bibliography{main}

@misc{yamada_cpprb_2019,
author = {Yamada, Hiroyuki},
month = {1},
title = {{cpprb}},
url = {https://gitlab.com/ymd_h/cpprb},
year = {2019}
}

@article{oecd_algorithms_2017,
  title={Algorithms and collusion: competition policy in the digital age},
  author={OECD},
  journal={OECD Roundtables on Competition Policy Papers},
  year={2017}
}

@InProceedings{schulman_trust_2015,
  title = 	 {Trust Region Policy Optimization},
  author = 	 {Schulman, John and Levine, Sergey and Abbeel, Pieter and Jordan, Michael and Moritz, Philipp},
  booktitle = 	 {Proceedings of the 32nd International Conference on Machine Learning},
  pages = 	 {1889--1897},
  year = 	 {2015},
  editor = 	 {Bach, Francis and Blei, David},
  volume = 	 {37},
  series = 	 {Proceedings of Machine Learning Research},
  address = 	 {Lille, France},
  month = 	 {07--09 Jul},
  publisher =    {PMLR},
  url = 	 {https://proceedings.mlr.press/v37/schulman15.html}
}

@article{abada_collusion_2024,
title = {Collusion by mistake: Does algorithmic sophistication drive supra-competitive profits?},
journal = {European Journal of Operational Research},
volume = {318},
number = {3},
pages = {927-953},
year = {2024},
issn = {0377-2217},
doi = {https://doi.org/10.1016/j.ejor.2024.06.006},
url = {https://www.sciencedirect.com/science/article/pii/S037722172400434X},
author = {Ibrahim Abada and Xavier Lambin and Nikolay Tchakarov}
}

@article{abreu_extremal_1986,
  title         = {Extremal equilibria of oligopolistic supergames},
  volume        = {39},
  issn          = {0022-0531},
  url           = {https://www.sciencedirect.com/science/article/pii/0022053186900256},
  doi           = {10.1016/0022-0531(86)90025-6},
  language      = {en},
  number        = {1},
  urldate       = {2022-06-26},
  journal       = {Journal of Economic Theory},
  author        = {Abreu, Dilip},
  month         = jun,
  year          = {1986},
  pages         = {191--225}
}

@inproceedings{adamczyk_average_2025,
  title         = {Average-Reward Soft Actor-Critic},
  author        = {Adamczyk, Jacob and Makarenko, Volodymyr and Tiomkin, Stas and Kulkarni, Rahul V},
  booktitle     = {Reinforcement Learning Conference},
  year          = {2025}
}

@article{asker_artificial_2022,
  title         = {Artificial Intelligence, Algorithm Design, and Pricing},
  volume        = {112},
  issn          = {2574-0768},
  url           = {https://www.aeaweb.org/articles?id=10.1257/pandp.20221059},
  doi           = {10.1257/pandp.20221059},
  language      = {en},
  urldate       = {2023-02-10},
  journal       = {AEA Papers and Proceedings},
  author        = {Asker, John and Fershtman, Chaim and Pakes, Ariel},
  month         = may,
  year          = {2022},
  pages         = {452--456}
}

@article{assad_algorithmic_2023,
  title         = {Algorithmic Pricing and Competition: Empirical Evidence from the German Retail Gasoline Market},
  issn          = {0022-3808, 1537-534X},
  shorttitle    = {Algorithmic Pricing and Competition},
  url           = {https://www.journals.uchicago.edu/doi/10.1086/726906},
  doi           = {10.1086/726906},
  language      = {en},
  urldate       = {2023-07-31},
  journal       = {Journal of Political Economy},
  author        = {Assad, Stephanie and Clark, Robert and Ershov, Daniel and Xu, Lei},
  month         = jul,
  year          = {2023},
  pages         = {726906}
}

@inproceedings{banchio_adaptive_2022,
  title         = {Adaptive Algorithms and Collusion via Coupling},
  author        = {Banchio, Martino and Mantegazza, Giacomo},
  booktitle     = {Proceedings of the 24th ACM Conference on Economics and Computation (EC '23)},
  year          = {2023},
  publisher     = {Association for Computing Machinery},
  address       = {London, United Kingdom},
  doi           = {10.1145/3580507.3597726}
}

@article{bergstra_random_2012,
  title         = {Random search for hyper-parameter optimization},
  volume        = {13},
  issn          = {1532-4435},
  number        = {null},
  journal       = {The Journal of Machine Learning Research},
  author        = {Bergstra, James and Bengio, Yoshua},
  month         = feb,
  year          = {2012},
  pages         = {281--305}
}

@article{borgers_learning_1997,
  title         = {Learning through reinforcement and replicator dynamics},
  author        = {B{\"o}rgers, Tilman and Sarin, Rajiv},
  journal       = {Journal of economic theory},
  volume        = {77},
  number        = {1},
  pages         = {1--14},
  year          = {1997},
  publisher     = {Elsevier}
}

@article{byrne_learning_2019,
  title         = {Learning to Coordinate: A Study in Retail Gasoline},
  volume        = {109},
  issn          = {0002-8282},
  shorttitle    = {Learning to Coordinate},
  url           = {http://www.aeaweb.org/articles?id=10.1257/aer.20170116},
  doi           = {10.1257/aer.20170116},
  language      = {en},
  number        = {2},
  urldate       = {2022-06-24},
  journal       = {American Economic Review},
  author        = {Byrne, David P. and de Roos, Nicolas},
  month         = feb,
  year          = {2019},
  pages         = {591--619}
}

@article{calvano_artificial_2020,
  title         = {Artificial Intelligence, Algorithmic Pricing, and Collusion},
  volume        = {110},
  issn          = {0002-8282},
  url           = {http://www.aeaweb.org/articles?id=10.1257/aer.20190623},
  doi           = {10.1257/aer.20190623},
  language      = {en},
  number        = {10},
  urldate       = {2022-04-18},
  journal       = {American Economic Review},
  author        = {Calvano, Emilio and Calzolari, Giacomo and Denicol\`{o}, Vincenzo and Pastorello, Sergio},
  month         = oct,
  year          = {2020},
  pages         = {3267--3297}
}

@article{calvano_protecting_2020,
  title         = {Protecting consumers from collusive prices due to AI},
  volume        = {370},
  issn          = {0036-8075, 1095-9203},
  url           = {https://www.science.org/doi/10.1126/science.abe3796},
  doi           = {10.1126/science.abe3796},
  language      = {en},
  number        = {6520},
  urldate       = {2023-07-31},
  journal       = {Science},
  author        = {Calvano, Emilio and Calzolari, Giacomo and Denicol\`{o}, Vincenzo and Harrington, Joseph E. and Pastorello, Sergio},
  month         = nov,
  year          = {2020},
  pages         = {1040--1042}
}

@article{calvano_algorithmic_2023,
  title         = {Algorithmic collusion: Genuine or spurious?},
  author        = {Calvano, Emilio and Calzolari, Giacomo and Denicol{\`o}, Vincenzo and Pastorello, Sergio},
  journal       = {International Journal of Industrial Organization},
  pages         = {102973},
  year          = {2023},
  publisher     = {Elsevier}
}

@book{cesa_prediction_2006,
  title         = {Prediction, learning, and games},
  author        = {Cesa-Bianchi, Nicolo and Lugosi, G{\'a}bor},
  year          = {2006},
  publisher     = {Cambridge university press}
}

@article{cesa-bianchi_boltzmann_2017,
  title         = {Boltzmann exploration done right},
  author        = {Cesa-Bianchi, Nicol{\`o} and Gentile, Claudio and Lugosi, G{\'a}bor and Neu, Gergely},
  journal       = {Advances in neural information processing systems},
  volume        = {30},
  year          = {2017}
}

@article{crandall_cooperating_2018,
  title         = {Cooperating with machines},
  volume        = {9},
  number        = {1},
  journal       = {Nature communications},
  author        = {Crandall, Jacob W and Oudah, Mayada and Ishowo-Oloko, Fatimah and Abdallah, Sherief and Bonnefon, Jean-Fran\c{c}ois and Cebrian, Manuel and Shariff, Azim and Goodrich, Michael A and Rahwan, Iyad and {others}},
  year          = {2018},
  note          = {Publisher: Nature Publishing Group},
  pages         = {1--12}
}

@article{crandall_towards_2014,
  title         = {Towards minimizing disappointment in repeated games},
  volume        = {49},
  journal       = {Journal of Artificial Intelligence Research},
  author        = {Crandall, Jacob W},
  year          = {2014},
  pages         = {111--142}
}

@article{den_artificial_2022,
  title         = {Artificial collusion: Examining supracompetitive pricing by Q-learning algorithms},
  author        = {den Boer, Arnoud V and Meylahn, Janusz M and Schinkel, Maarten Pieter},
  journal       = {Amsterdam Law School Research Paper},
  number        = {2022-25},
  year          = {2022}
}

@inproceedings{dudik_maximum_2006,
  title         = {Maximum entropy distribution estimation with generalized regularization},
  author        = {Dud{\'\i}k, Miroslav and Schapire, Robert E},
  booktitle     = {International Conference on Computational Learning Theory},
  pages         = {123--138},
  year          = {2006},
  organization  = {Springer}
}

@misc{foster_statistical_2021,
  author        = {Foster, Dylan},
  title         = {The Statistical Complexity of Interactive Decision Making},
  year          = {2022},
  month         = {October},
  day           = {11},
  howpublished  = {Talk at the Simons Institute for the Theory of Computing},
  note          = {Workshop on Structure of Constraints in Sequential Decision-Making},
  organization  = {Simons Institute for the Theory of Computing}
}

@inproceedings{haarnoja_soft_2018,
  title         = {Soft Actor-Critic: Off-Policy Maximum Entropy Deep Reinforcement Learning with a Stochastic Actor},
  shorttitle    = {Soft Actor-Critic},
  url           = {https://proceedings.mlr.press/v80/haarnoja18b.html},
  language      = {en},
  urldate       = {2022-04-17},
  booktitle     = {Proceedings of the 35th International Conference on Machine Learning},
  publisher     = {PMLR},
  author        = {Haarnoja, Tuomas and Zhou, Aurick and Abbeel, Pieter and Levine, Sergey},
  month         = jul,
  year          = {2018},
  note          = {ISSN: 2640-3498},
  pages         = {1861--1870}
}

@article{henderson_deep_2018,
  title         = {Deep Reinforcement Learning That Matters},
  volume        = {32},
  url           = {https://ojs.aaai.org/index.php/AAAI/article/view/11694},
  doi           = {10.1609/aaai.v32i1.11694},
  number        = {1},
  journal       = {Proceedings of the AAAI Conference on Artificial Intelligence},
  author        = {Henderson, Peter and Islam, Riashat and Bachman, Philip and Pineau, Joelle and Precup, Doina and Meger, David},
  year          = {2018},
  month         = {Apr.}
}

@article{hettich_algorithmic_2021,
  title         = {Algorithmic Collusion: Insights from Deep Learning},
  author        = {Hettich, Matthias},
  year          = {2021},
  doi           = {10.2139/ssrn.3785966},
  journal       = {SSRN}
}

@article{hornik_multilayer_1989,
  title         = {Multilayer feedforward networks are universal approximators},
  volume        = {2},
  number        = {5},
  journal       = {Neural networks},
  author        = {Hornik, Kurt and Stinchcombe, Maxwell and White, Halbert},
  year          = {1989},
  note          = {Publisher: Elsevier},
  pages         = {359--366}
}

@article{ismailov_three_2023,
  title         = {A three layer neural network can represent any multivariate function},
  author        = {Ismailov, Vugar E},
  journal       = {Journal of Mathematical Analysis and Applications},
  volume        = {523},
  number        = {1},
  pages         = {127096},
  year          = {2023},
  publisher     = {Elsevier}
}

@article{johnson_platform_2023,
  title         = {Platform design when sellers use pricing algorithms},
  author        = {Johnson, Justin P and Rhodes, Andrew and Wildenbeest, Matthijs},
  journal       = {Econometrica},
  volume        = {91},
  number        = {5},
  pages         = {1841--1879},
  year          = {2023},
  publisher     = {Wiley Online Library}
}

@article{kianercy_dynamics_2012,
  title         = {Dynamics of Boltzmann Q learning in two-player two-action games},
  author        = {Kianercy, Ardeshir and Galstyan, Aram},
  journal       = {Physical Review E--Statistical, Nonlinear, and Soft Matter Physics},
  volume        = {85},
  number        = {4},
  pages         = {041145},
  year          = {2012},
  publisher     = {APS}
}

@inproceedings{kingma_dp_adam_2015,
  title         = {Adam: A Method for Stochastic Optimization},
  shorttitle    = {Adam},
  url           = {https://dare.uva.nl/personal/pure/en/publications/adam-a-method-for-stochastic-optimization(a20791d3-1aff-464a-8544-268383c33a75).html},
  language      = {en},
  urldate       = {2022-06-27},
  booktitle     = {International Conference on Learning Representations (ICLR)},
  author        = {Kingma, D.P. and Ba, L.J. and {Amsterdam Machine Learning lab (IVI, FNWI)}},
  year          = {2015}
}

@article{kiran_deep_2021,
  title         = {Deep reinforcement learning for autonomous driving: A survey},
  author        = {Kiran, B Ravi and Sobh, Ibrahim and Talpaert, Victor and Mannion, Patrick and Al Sallab, Ahmad A and Yogamani, Senthil and P{\'e}rez, Patrick},
  journal       = {IEEE Transactions on Intelligent Transportation Systems},
  volume        = {23},
  number        = {6},
  pages         = {4909--4926},
  year          = {2021},
  publisher     = {IEEE}
}

@article{klein_autonomous_2021,
  title         = {Autonomous algorithmic collusion: Q-learning under sequential pricing},
  volume        = {52},
  issn          = {1756-2171},
  shorttitle    = {Autonomous algorithmic collusion},
  url           = {http://onlinelibrary.wiley.com/doi/abs/10.1111/1756-2171.12383},
  doi           = {10.1111/1756-2171.12383},
  language      = {en},
  number        = {3},
  urldate       = {2022-06-24},
  journal       = {The RAND Journal of Economics},
  author        = {Klein, Timo},
  year          = {2021},
  pages         = {538--558}
}

@article{kuindersma_optimization-based_2015,
  title         = {Optimization-based locomotion planning, estimation, and control design for the atlas humanoid robot},
  copyright     = {Creative Commons Attribution-Noncommercial-Share Alike},
  issn          = {0929-5593},
  url           = {https://dspace.mit.edu/handle/1721.1/110533},
  language      = {en\_US},
  urldate       = {2022-06-19},
  journal       = {MIT web domain},
  author        = {Kuindersma, Scott and Deits, Robin and Fallon, Maurice and Valenzuela, Andr\'{e}s and Dai, Hongkai and Permenter, Frank and Koolen, Twan and Marion, Pat and Tedrake, Russ},
  month         = jul,
  year          = {2015},
  note          = {Accepted: 2017-07-07T15:29:29Z Publisher: Springer-Verlag}
}

@inproceedings{lillicrap_continuous_2021,
  title         = {Continuous control with deep reinforcement learning.},
  url           = {https://openreview.net/forum?id=tX_O8O-8Zl},
  language      = {en},
  urldate       = {2022-06-27},
  booktitle     = {ICLR (Poster)},
  author        = {Lillicrap, Timothy P. and Hunt, Jonathan J. and Pritzel, Alexander and Heess, Nicolas and Erez, Tom and Tassa, Yuval and Silver, David and Wierstra, Daan},
  month         = jan,
  year          = {2021}
}

@article{lin_self-improving_1992,
  title         = {Self-improving reactive agents based on reinforcement learning, planning and teaching},
  volume        = {8},
  issn          = {1573-0565},
  url           = {https://doi.org/10.1007/BF00992699},
  doi           = {10.1007/BF00992699},
  language      = {en},
  number        = {3},
  urldate       = {2022-04-20},
  journal       = {Machine Learning},
  author        = {Lin, Long-Ji},
  month         = may,
  year          = {1992},
  pages         = {293--321}
}

@article{liu_sample_2022,
  title         = {Sample-efficient reinforcement learning of partially observable markov games},
  author        = {Liu, Qinghua and Szepesv{\'a}ri, Csaba and Jin, Chi},
  journal       = {Advances in Neural Information Processing Systems},
  volume        = {35},
  pages         = {18296--18308},
  year          = {2022}
}

@inproceedings{mahadevan_optimality_1996,
  title         = {Optimality criteria in reinforcement learning},
  booktitle     = {Proceedings of the AAAI Fall Symposium on Learning Complex Behaviors in Adaptive Intelligent Systems},
  publisher     = {Citeseer},
  author        = {Mahadevan, Sridhar},
  year          = {1996}
}

@article{mnih_human-level_2015,
  title         = {Human-level control through deep reinforcement learning},
  volume        = {518},
  copyright     = {2015 Nature Publishing Group, a division of Macmillan Publishers Limited. All Rights Reserved.},
  issn          = {1476-4687},
  url           = {https://www.nature.com/articles/nature14236},
  doi           = {10.1038/nature14236},
  language      = {en},
  number        = {7540},
  urldate       = {2022-04-18},
  journal       = {Nature},
  author        = {Mnih, Volodymyr and Kavukcuoglu, Koray and Silver, David and Rusu, Andrei A. and Veness, Joel and Bellemare, Marc G. and Graves, Alex and Riedmiller, Martin and Fidjeland, Andreas K. and Ostrovski, Georg and Petersen, Stig and Beattie, Charles and Sadik, Amir and Antonoglou, Ioannis and King, Helen and Kumaran, Dharshan and Wierstra, Daan and Legg, Shane and Hassabis, Demis},
  month         = feb,
  year          = {2015},
  note          = {Number: 7540 Publisher: Nature Publishing Group},
  pages         = {529--533}
}

@article{moravcik_deepstack_2017,
  title         = {DeepStack: Expert-level artificial intelligence in heads-up no-limit poker},
  volume        = {356},
  shorttitle    = {DeepStack},
  url           = {https://www.science.org/doi/full/10.1126/science.aam6960},
  doi           = {10.1126/science.aam6960},
  number        = {6337},
  urldate       = {2022-06-28},
  journal       = {Science},
  author        = {Morav\v{c}\'{\i}k, Matej and Schmid, Martin and Burch, Neil and Lis\'{y}, Viliam and Morrill, Dustin and Bard, Nolan and Davis, Trevor and Waugh, Kevin and Johanson, Michael and Bowling, Michael},
  month         = may,
  year          = {2017},
  note          = {Publisher: American Association for the Advancement of Science},
  pages         = {508--513}
}

@inproceedings{naik_discounted_2019,
  title         = {Discounted Reinforcement Learning is Not an Optimization Problem},
  booktitle     = {NeurIPS 2019 Optimization Foundations for Reinforcement Learning Workshop},
  author        = {Naik, Abhishek and Shariff, Roshan and Yasui, Niko and Sutton, Richard S.},
  year          = {2019}
}

@inproceedings{neu_unified_2017,
  author        = {Gergely Neu and Vicen{\c{c}} G{\'{o}}mez and Anders Jonsson},
  title         = {A unified view of entropy-regularized Markov decision processes},
  booktitle     = {NeurIPS Workshop: Deep Reinforcement Learning Symposium},
  year          = {2017},
  eprint        = {1705.07798}
}

@inproceedings{nota_is_2020,
  title         = {Is the Policy Gradient a Gradient?},
  doi           = {10.5555/3398761.3398871},
  booktitle     = {AAMAS},
  author        = {Nota, Chris and Thomas, P.},
  year          = {2020}
}

@article{possnig_learning_2023,
  title         = {Learning to Best Reply: On the Consistency of Multi-Agent Reinforcement Learning},
  author        = {Possnig, Clemens},
  journal       = {Working paper},
  year          = {2023}
}

@article{possnig_reinforcement_2023,
  title         = {Reinforcement learning and collusion},
  author        = {Possnig, Clemens},
  journal       = {Working paper},
  year          = {2023}
}

@article{rubinstein_equilibrium_1979,
  title         = {Equilibrium in supergames with the overtaking criterion},
  volume        = {21},
  issn          = {00220531},
  url           = {https://linkinghub.elsevier.com/retrieve/pii/0022053179900024},
  doi           = {10.1016/0022-0531(79)90002-4},
  language      = {en},
  number        = {1},
  urldate       = {2022-08-01},
  journal       = {Journal of Economic Theory},
  author        = {Rubinstein, Ariel},
  month         = aug,
  year          = {1979},
  pages         = {1--9}
}

@article{schwalbe_algorithms_2018,
  title         = {Algorithms, Machine Learning and Collusion},
  volume        = {14},
  issn          = {1744-6414},
  url           = {https://doi.org/10.1093/joclec/nhz004},
  doi           = {10.1093/joclec/nhz004},
  number        = {4},
  urldate       = {2022-03-28},
  journal       = {Journal of Competition Law \& Economics},
  author        = {Schwalbe, Ulrich},
  month         = dec,
  year          = {2018},
  pages         = {568--607}
}

@incollection{singh_learning_1994,
  title         = {Learning without state-estimation in partially observable Markovian decision processes},
  author        = {Singh, Satinder P and Jaakkola, Tommi and Jordan, Michael I},
  booktitle     = {Machine Learning Proceedings 1994},
  pages         = {284--292},
  year          = {1994},
  publisher     = {Elsevier}
}

@book{sutton_reinforcement_2018,
  title         = {Reinforcement Learning: An Introduction},
  isbn          = {978-0-262-35270-3},
  shorttitle    = {Reinforcement Learning, second edition},
  language      = {en},
  publisher     = {MIT Press},
  author        = {Sutton, Richard S. and Barto, Andrew G.},
  month         = nov,
  year          = {2018},
  note          = {Google-Books-ID: uWV0DwAAQBAJ}
}

@inproceedings{tassa_synthesis_2012,
  title         = {Synthesis and stabilization of complex behaviors through online trajectory optimization},
  doi           = {10.1109/IROS.2012.6386025},
  booktitle     = {2012 IEEE/RSJ International Conference on Intelligent Robots and Systems},
  author        = {Tassa, Yuval and Erez, Tom and Todorov, Emanuel},
  month         = oct,
  year          = {2012},
  note          = {ISSN: 2153-0866},
  pages         = {4906--4913}
}

@misc{uscongress_preventing_2024,
  title         = {Preventing Algorithmic Collusion Act - S.3686 - 118th Congress},
  author        = {{U.S. Congress}},
  year          = {2024},
  howpublished  = {https://www.congress.gov/bill/118th-congress/senate-bill/3686}
}

@inproceedings{van_hasselt_double_2010,
  title         = {Double Q-learning},
  volume        = {23},
  url           = {https://proceedings.neurips.cc/paper/2010/hash/091d584fced301b442654dd8c23b3fc9-Abstract.html},
  urldate       = {2022-04-18},
  booktitle     = {Advances in Neural Information Processing Systems},
  publisher     = {Curran Associates, Inc.},
  author        = {van Hasselt, Hado},
  year          = {2010}
}

@incollection{van_hasselt_reinforcement_2012,
  address       = {Berlin, Heidelberg},
  series        = {Adaptation, {Learning}, and {Optimization}},
  title         = {Reinforcement Learning in Continuous State and Action Spaces},
  isbn          = {978-3-642-27645-3},
  url           = {https://doi.org/10.1007/978-3-642-27645-3_7},
  language      = {en},
  urldate       = {2022-04-19},
  booktitle     = {Reinforcement Learning: State-of-the-Art},
  publisher     = {Springer},
  author        = {van Hasselt, Hado},
  editor        = {Wiering, Marco and van Otterlo, Martijn},
  year          = {2012},
  doi           = {10.1007/978-3-642-27645-3_7},
  pages         = {207--251}
}

@inproceedings{yang_mean_2019,
  title         = {A Mean Field Theory of Batch Normalization},
  booktitle     = {Google Research},
  author        = {Yang, Greg and Sohl-dickstein, Jascha and Pennington, Jeffrey and Schoenholz, Sam and Rao, Vinay},
  year          = {2019}
}

@inproceedings{yu_towards_2018,
  title         = {Towards Sample Efficient Reinforcement Learning.},
  author        = {Yu, Yang},
  booktitle     = {IJCAI},
  pages         = {5739--5743},
  year          = {2018}
}

\clearpage

\appendix
\titleformat{\section}
  {\normalfont\Large\bfseries} % Format of the title font
  {Appendix \thesection:}      % Label: "Appendix A:"
  {1em}                        % Space between label and title
  {}                           % Code before the title body
\section{The average-reward objective}
\label{appendix:avgrew}
The ordering of policies is partial \citep{naik_discounted_2019} under function approximation for infinite-horizon tasks.
That is, if there is no natural ``ending'' to a task, the definition given above for the optimal policy (maximising the sum of discounted rewards) does not allow for comparison between any two pairs of policies: some policies may achieve higher rewards in some state and lower in others.
In principle, this is not an issue with Q-learning because it can represent potentially any policy and it can be proven that there exists one that maximises value at every state simultaneously.

However, this is not feasible under function approximation, and even more so given a continuous state variable.
The space of possible policies is so large that the optimal policy is not representable, so the objective becomes to find the best representable policy.

\citet{naik_discounted_2019} show that in the case of methods requiring function approximation, finding the best representable policy in terms of discounted reward is not a well-defined optimisation problem.
They suggest that a metric that can be maximised with stochastic gradient ascent is average reward.
It can be defined if the Markov decision process is ergodic, that is for any policy $\sigma(a|s)$ there exists a steady-state distribution of states $\mu_\sigma(s) = \lim\limits_{t \rightarrow \infty} \Pr[S_t = s |(A_0, ..., A_{t-1} \sim \sigma)]$ that is independent of the initial state $S_0$.

Average reward is of course unknown, but it can be estimated by keeping a running average that is initially equal to 0 and is then updated as follows:
\begin{equation}
  \hat{\pi}_{t+1}(\sigma) = (1 - \lambda_R) \hat{\pi}_t(\sigma) + \lambda_R[\pi(s, a, s') + \hat{v}(s') - \hat{v}(s)]
  \label{eq:avgrewupd}
\end{equation}
or, using the action-value function instead,
\begin{equation}
  \hat{\pi}_{t+1}(\sigma) = (1 - \lambda_R) \hat{\pi}_t(\sigma) + \lambda_R[\pi(s, a, s') + \hat{q}(s', a') - \hat{q}(s, a)]
  \label{eq:avgrewupdq}
\end{equation}
where $\lambda_R \in (0, 1)$ is a learning rate, $\hat{v}(s)$ and $\hat{q}(s, a)$ are estimations of the value and action-value functions, $S_{t +1} = s', S_t = s, A_t = a, A_{t + 1} = a'$, and $a' \sim \sigma$ is given by the current policy estimation.

These formulas do not merely represent a running average of rewards but include the temporal-difference error $\varepsilon = \pi(S_t, A_t) - \hat{\pi}_t + \hat{q}(S_{t + 1}, A_{t + 1}) - \hat{q}(S_t, A_t)$.
The reason for this is that it ensures convergence: by updating the estimate as $\hat{\pi}_{t + 1} = \hat{\pi}_t + d \varepsilon$, once action-value function estimation converges we get, using $\pi_t$ as a shorthand for $\pi(S_t, A_t)$:
\begin{align}
\nonumber \varepsilon &= \pi_t - \hat{\pi}_t + \hat{q}(S_{t + 1}, A_{t + 1}) - \hat{q}(S_t, A_t) \\
\nonumber &= \pi_t - \hat{\pi}_t + (\pi_{t + 1} - \bar{\pi}(\sigma^*) + \pi_{t + 2} - \bar{\pi}(\sigma^*) + ...) - (\pi_t - \bar{\pi}(\sigma^*) + \pi_{t + 1} - \bar{\pi}(\sigma^*) + ...) \\
\nonumber &= \pi_t - \hat{\pi}_t - (\pi_t - \bar{\pi}(\sigma^*)) \\
\nonumber &= \bar{\pi}(\sigma^*) - \hat{\pi}_t.
\end{align}
Therefore, the estimation converges to the average reward under the optimal policy, as it is being updated as $\hat{\pi}_{t + 1} = (1 - \lambda_R) \hat{\pi}_t + \lambda_R \bar{\pi}(\sigma^*)$.

I refer to the policy maximising the objective function $\bar{\pi}(\sigma)$ as the optimal policy, even though under function approximation it is technically the optimal representable policy.
Most policy gradient methods proposed in the literature mention the discounted return formulation of the policy gradient theorem in their formulation, but actually update their parameters using estimates that do not approximate the performance measure gradient \citep{nota_is_2020}, possibly leading to sub-optimal results.
This issue does not concern my work because I am optimising in the average-reward formulation and I do not use a discount factor for the rewards or the state distribution.

\section{Technical specifications and hyperparameters}
\label{appendix:specs}

Following the reinforcement learning literature, I choose to fix a probability distribution and learn its parameters instead of using even more sophisticated non-parametric policy estimation methods.
While learning to approximate $\sigma(a|s)$ nonparametrically might be feasible in principle, in practice it is much more tractable to either learn a deterministic mapping of spaces to actions, the approach used in deep deterministic policy gradient \citep{lillicrap_continuous_2021}, or learn the parameters of a given distribution.
An agent using soft actor-critic chooses its actions by sampling from a normal distribution whose mean and variance are state-dependent and parameterised by the parameter vector $\boldsymbol{\theta}$, i.e.
\begin{equation}
  A_t \sim \mathcal{N}(\mu(s; \boldsymbol{\theta}), \Sigma^2(s; \boldsymbol{\theta}))
\end{equation}
where $\mu, \Sigma: \mathcal{S} \rightarrow \mR$ are the outputs of the actor network.
The parameter vectors $\boldsymbol{\theta}$ and $\boldsymbol{w}$ thus represent the weights and biases of networks trained via stochastic gradient descent and backpropagation to optimise the objective functions stated in the main text.
Inputs to the networks are batches of size $b$ containing tuples $(S_t, A_t, \pi_t, S_{t+1})$ randomly sampled from an array containing the last $d$ experiences, again using $\pi_t$ as a shorthand for $\pi(S_t, A_t)$.
To fill the buffer, I initially force agents to play entirely random prices for a number of time steps equal to the batch size, as is standard practice in reinforcement learning.
The expectations are taken over a batch of samples from a buffer, using an additional network with lagged updates to reduce estimation variance, as explained further down.

Actions sampled from a normal distribution are unbounded, as the density function of a normal distribution is never zero, but it is often desirable (like in our case) to constrain actions inside some interval.
To do so, \citet{haarnoja_soft_2018} propose applying a squashing function whose range is a bounded interval to the normally-distributed network output, such as the hyperbolic tangent function whose range is $[-1, 1]$.
The formulation of the probability density function of the resulting distribution is shown in the appendices of \citet{haarnoja_soft_2018}.

Convergence proofs for stochastic gradient descent require that the samples in each batch be independent and identically distributed, so that the estimator obtained equals the actual function gradient in expectation.
However, this assumption is violated when optimising on subsequent experiences such as what happens in temporal-difference learning, as any two subsequent tuples $(S_t,A_t,\pi_t,S_{t+1})$ and $(S_{t+1},A_{t+1},\pi_{t+1},S_{t+2})$ are correlated with one another and share one element, the middle state $S_{t+1}$.

Replay memory, also known as experience replay, was first introduced by \citet{lin_self-improving_1992} to solve this issue.
This method stores the agent's experience at each time step in a replay buffer containing tuples $(S_t,A_t,\pi_t,S_{t+1})$.
A buffer usually has a maximum size $d$ and older experiences are removed and substituted with newer experiences in a first-in-first-out fashion once the buffer is full.
In my implementation, the replay buffer is implemented in C++ for performance using the cpprb library \citep{yamada_cpprb_2019}.

At each time step, having a replay memory allows for performing multiple gradient ascent steps based on a batch of experiences sampled uniformly at random from the replay buffer.

Instead of $S_{t+1}$ becoming the new state for the next update as it would in the usual form of temporal-difference learning, a new unconnected experience is drawn from the replay memory to supply data for the next update.
Off-policy algorithms do not need to be applied along connected trajectories, and the possibility to use each stored experience for many updates allows for more efficient learning from experience.

Experience replay reduces the variance of the updates because successive updates are not correlated with one another as they would be if gradient descent was performed on subsequent experience tuples where the last element of the first tuple is always the first element of the second tuple.
By removing the dependence of successive experiences on the current weights, experience replay eliminates one source of instability.

Moreover, experience replay allows for more efficient parallel computation, for example by using graphics processing units (GPUs) as is standard practice in supervised machine learning.

Replay buffers have been conceived with Markov decision processes in mind.
In a multi-agent environment where all agents are learning, taking an action in a given state many periods ago probably did not give out the same reward as it does now because the agent is still learning, which makes the buffer size parameter a proxy for the degree of ``smoothing the non-stationarity'' of the environment, at least in principle, by reducing the variance of the updates.

In addition to the content of the experiences themselves, another source of correlated output is bootstrapping itself.
The objective function in eq. \ref{eq:sacqloss} defines the optimisation problem that gives an estimate of the action-value function, but uses the estimated action-value function itself to compute the value function of the future state that yields the optimisation target (the second term inside the expectation) and the target is thus correlated with the estimand.
Again, this translates into a correlation between inputs and impedes convergence.
This is in contrast, for example, to what happens in supervised learning, or indeed in soft actor-critic with its regularisation, where the optimisation target is an exogenously fixed number that does not depend on the current estimation.
To solve this other issue, \citet{mnih_human-level_2015} propose computing the TD error using a separate network, known as the target network, parameterised by $\boldsymbol{u}$, with $\boldsymbol{u}_0 = \boldsymbol{w}_0$ at $t = 0$.
$\boldsymbol{u}$ is then updated as a moving average between $\boldsymbol{u}$ and $\boldsymbol{w}$, i.e. $\boldsymbol{u}_{t + 1} = \tau \boldsymbol{u}_t + (1 - \tau) \boldsymbol{w}_{t+1}$, where $\tau$ is a learning rate.
This detaches the optimisation target from the current estimation, thus allowing for smoother convergence, and it only introduces a very slight lag in action-value function estimation if $\tau$ is small enough.
Usually, this hyperparameter is in the order of 0.001, but models are not overly sensitive to it.
As an alternative, \citet{mnih_human-level_2015} also explore setting $\boldsymbol{u} = \boldsymbol{w}$ every few periods, observing very similar results.

In the average-reward formulation, the TD error has to be computed with the target network to make average reward estimation consistent (proof that eq. \ref{eq:avgrewupdq} converges to average reward is provided in Appendix \ref{appendix:avgrew}).

Constructing a reinforcement learning model implies tuning many implicit hyperparameters in the form of architectural choices, and since the release of the first article on soft actor-critic \citep{haarnoja_soft_2018} various improvements have been proposed that are both specific to soft-actor critic and generally applicable to machine learning architectures.
Machine learning is a very active research field, but models in deep reinforcement learning do not necessarily benefit from architectural improvements that have been developed for supervised learning.
For example, batch normalisation has been found to induce gradient explosion \citep{yang_mean_2019}, which hinders convergence in reinforcement learning.
Temporal-difference methods are not exempt from this ``architecture tuning'', as they do not only require tuning of the learning and exploration rates but entail a choice of an exploration strategy, as well as between on- or off-policy learning and whether or not to implement bias/variance reduction techniques such as $n$-step return or double Q-learning.

The TD error may be formulated in terms of an action-value function estimation, so I decided not to employ a separate value network and compute the TD error as $\varepsilon = \pi(s, a, s') - \bar{\pi}_t(\sigma) + \hat{q}(s', a') - \hat{q}(s, a)$ \citep{haarnoja_soft_2018}.
I also use orthogonal initialisation for the hidden-layer weights of both the actor and critic networks; the actor output layers use a small uniform initialisation (gain $=0.1$) with the log-standard-deviation bias set to 1.0, and the critic output layer uses orthogonal weights (gain $=0.01$) with a bias of 10.0.
This optimistic initialisation is intended to encourage exploration, following the discussion in Section 2.6 of \citet{sutton_reinforcement_2018}.

I use neural networks with two hidden layers with leaky ReLU as their activation function and employ two Q-networks to reduce overestimation of the action-value function \citep{van_hasselt_double_2010}. The estimate $\hat{q}(s, a)$ is given by the mean of the two estimates.

Agents do not discount rewards, and they keep a running estimate of average reward as in eq. \ref{eq:avgrewupdq}, using action-value estimates computed using the target network.

The critic network has an input layer of size $\dim(\mathcal{A}) + \dim(\mathcal{S}) = n + 1$.
The input layer is followed by two hidden layers of size $m_C$ with leaky ReLU activation functions and an output layer of size 1.
The actor network has the same structure, but with hidden layers of size $m_A$, an input layer of size $\dim(\mathcal{S}) = n$, and an output layer of size 2 (mean and standard deviation of a normal distribution).

I use the Adam stochastic optimiser \citep{kingma_dp_adam_2015} to perform gradient descent, with a constant step size $\lambda_A$ for the actor, $\lambda_C$ for the critic, and $\lambda_T$ for the temperature.
I perform updates at every step, sampling batches of size $b$ from a replay buffer of maximum length $d$.
For the first $b$ steps, I let agents play entirely random actions.
These architectural choices lead to a relatively simple model and have been empirically found to improve convergence, even though at the moment there is no formal explanation for their effectiveness.

An exhaustive grid search is not feasible due to the runtime of a single experiment, so I ran a random sweep over the parameter space, which has been shown to perform better than grid search \citep{bergstra_random_2012} in finding good hyperparameters for neural networks.
Reported results have been obtained using the hyperparameters specified in Table \ref{table:hyppar}, but they are robust to changes in these hyperparameters that do not exceed an order of magnitude.

I ran experiments on Ubuntu 25.04 running on a workstation equipped with an AMD Ryzen Threadripper 7960X, two NVIDIA GeForce RTX 4090 graphics cards, and 512 GiB of RAM.

\begin{table}[t]
\caption{Experiment Hyperparameters}
\label{table:hyppar}
\centering
\begin{tabular}{|l|l|l|}
\hline
\textbf{Symbol} & \textbf{Hyperparameter} & \textbf{Value} \\ \hline
$\bar{h} $& Target entropy & -1.0\\ \hline
$m_C$ & Hidden layer size (critic network) & 256       \\ \hline
$m_A$ & Hidden layer size (actor network) & 1024       \\ \hline
$L_C$ & Hidden layers (critic) & 2 \\ \hline
$L_A$ & Hidden layers (actor) & 2 \\ \hline
$\phi$ & Hidden activation & \texttt{leaky\_relu} \\ \hline
$b$ & Batch size &  128       \\ \hline
$d$ & Replay buffer size &  100{,}000       \\ \hline
$\tau$ & Target network update rate &  0.001       \\ \hline
$\lambda_A$ & Actor optimiser learning rate & 0.03 \\ \hline
$\lambda_C$ & Critic optimiser learning rate & 0.003 \\ \hline
$\lambda_T$ & Temperature optimiser learning rate & 0.003 \\ \hline
$\lambda_R$ & Average reward estimation learning rate & $0.01$ \\ \hline
\end{tabular}

\end{table}

\end{document}